\newif\ifsubmit\submitfalse
\newif\ifrevtex\revtextrue
\newif\ifonecol\onecoltrue

\ifrevtex
\ifsubmit                       
\documentstyle[epsf,journals_aps,jprdefs,prd,aps,preprint]{revtex}
\draft
\else                           
\documentstyle[epsf,journals_aps,jprdefs,times,aps,prd,apsjpr,floats,twocolumn]{revtex}
\onecolfalse
\draft

\fi
\renewcommand{\gsim}{\gtrsim}   
\renewcommand{\lsim}{\lesssim}
\newcommand{\capsize}{\relax}
\else                           
\documentstyle[11pt,a4,journals_aps,jprdefs,epsf]{article}
\newcommand{\capsize}{\small}

\setheadmargin{1.5in}{0pt}{0pt}
\fi

\begin{document}

\title{On the cosmic ray bound for models of extragalactic neutrino
production} 

\ifrevtex
\author{Karl Mannheim} 
\address{Universit{\"a}ts-Sternwarte,
Geismarlandstr.~11, D-37083 G{\"o}ttingen, Germany; kmannhe@uni-sw.gwdg.de}

\author{R.J. Protheroe} 
\address{Department of Physics and Mathematical
Physics, The University of Adelaide, Adelaide, Australia 5005;
rprother@physics.adelaide.edu.au} 

\author{J{\"o}rg P. Rachen}
\address{Sterrenkundig Instituut, Universiteit Utrecht, NL-3508 TA Utrecht,
The Netherlands; J.P.Rachen@astro.uu.nl} 

\else

\author{
{\sc Karl Mannheim}\\[3pt] 
Universit{\"a}ts-Sternwarte\\
Geismarlandstr.~11, D-37083 G{\"o}ttingen, Germany\\
{\it kmannhe@uni-sw.gwdg.de}\\[1pc]
{\sc R.J. Protheroe}\\[3pt] 
Department of Physics and Mathematical Physics\\
The University of Adelaide, Adelaide, Australia 5005\\
{\it rprother@physics.adelaide.edu.au}\\[1pc]
{\sc J{\"o}rg P. Rachen}\\[3pt]
Sterrenkundig Instituut\\
Universiteit Utrecht, NL-3508 TA Utrecht, The Netherlands\\
{\it J.P.Rachen@astro.uu.nl}
} 

\fi

\date{submitted to Phys.~Rev.~D: 22 December 1998; accepted: 26 June 2000} 

\ifonecol\maketitle\fi

\def\abstracttext{
We obtain the maximum diffuse neutrino intensity
predicted by hadronic photoproduction models of the type which have been
applied to the jets of active galactic nuclei or gamma ray bursts.  For
this, we compare the proton and gamma ray fluxes associated with
hadronic photoproduction in extragalactic neutrino sources with the
present experimental upper limit on cosmic ray protons and the
extragalactic gamma ray background, employing a transport calculation of
energetic protons traversing cosmic photon backgrounds.  We take into
account the effects of the photon spectral shape in the sources on the
photoproduction process, cosmological source evolution, the optical
depth for cosmic ray ejection, and discuss the possible effects of
magnetic fields in the vicinity of the sources.  
For photohadronic neutrino sources which are optically thin to the emission
of neutrons we find that the cosmic ray flux imposes a stronger bound
than the extragalactic gamma ray background in the energy range between
$10^5\GeV$ and $10^{11}\GeV$, as previously noted by Waxman \& Bahcall
\cite{WB99prd}.  We also determine the maximum contribution from the jets
of active galactic nuclei, using constraints set to their neutron opacity
by gamma-ray observations.  This present upper limit is consistent with the jets
of active galactic nuclei producing the extragalactic gamma ray background
hadronically, but we point out future observations in the GeV-to-TeV regime
could lower this limit.   We also
briefly discuss the contribution of gamma ray bursts to ultra-high
energy cosmic rays as it can be inferred from possible observations or
limits on their correlated neutrino fluxes. 
}

\ifonecol
\begin{abstract}
\abstracttext
\end{abstract}
\else
\abstract{\parindent1em
\abstracttext}
\fi

\ifrevtex
\fi

\ifonecol\relax\else\maketitle\fi

\section{Introduction}

The connection between the emission of cosmic rays, gamma rays, and neutrinos
from astrophysical accelerators is of considerable interest for the solution
of the problem of the origin of cosmic rays \cite{Berezinsky}.  The reason why a
fundamental relation between these components must exist can be understood as
follows.  Particle acceleration mechanisms in cosmic plasmas generally
require the presence of a magnetic field which is able to confine the
accelerated charged particles, i.e., electrons and protons (or ions).  The
accelerated electrons lose their energy quickly in synchrotron radiation in
the magnetic field.  These synchrotron photons provide a target for
accelerated protons to undergo photohadronic interactions, resulting in the
production of mesons, which decay.  The particles which eventually emerge
from this process are high energy photons, electrons (pairs), neutrons and
neutrinos.  Neutrinos are directly ejected due to their low interaction cross
section.  Gamma rays and secondary electrons initiate electromagnetic
cascades, shifting the power from ultra-high energies to energies below which
the absorption of gamma rays by pair production is unimportant
\cite{MKB91aa}.  Finally, the neutrons, which unlike the protons, are not
confined in the magnetic field, can escape and convert into cosmic ray
protons after $\beta$-decay, but their flux may be diminished by photoinduced
$n\to p$ conversions.  The branching ratios which distribute the available
energy into the different channels are thereby generally of order unity.
This leads to the conclusion that, {\em cosmic proton accelerators produce
cosmic rays, gamma-rays and neutrinos with comparable luminosities}
\cite{Man93aa}.

The fundamental relation between cosmic ray and gamma ray production has the
obvious consequence that AGN, which are known to produce a large fraction of
the gamma-rays in the Universe, are a prime candidate for the sources of
ultra-high energy cosmic rays (UHECR) \cite{Man93aa}.  The spectra of the
emitted GeV-TeV gamma radiation of AGN also agree with the predictions of a
hadronic production of these gamma rays \cite{Man98sci}.  Other prominent
gamma-ray sources, in particular the violent events connected with gamma-ray
bursts (GRB), have also been suggested as UHECR source candidates.  Moreover,
most of the extragalactic gamma-ray energy is found in a diffuse background
rather than in point sources, which allows for the possibility that the UHECR
sources could be relatively large objects which would have a low gamma-ray
surface brightness, such as radio galaxies \cite{RB93aa}, galaxy clusters
\cite{KRJ96apj,KRB97mnras}, or even larger structures \cite{NMA95apj}.  Whatever
the sources are, the fundamental relation between gamma-ray and neutrino fluxes
implies that, if in fact the extragalactic gamma-ray emission is due to hadronic
processes, a neutrino flux of a similar bolometric luminosity must exist.  This
prediction is the major motivation for high energy neutrino experiments, which
are currently operated, under construction, or planned.  In fact, most model
predictions for extragalactic high-energy neutrino fluxes have been made by
using the source model to determine the spectral shape, and then by normalizing
the total flux to some fraction of the diffuse extragalactic gamma-ray
background (EGRB) \cite{Man95app,HZ97apj}.

In contrast to the limits set by gamma ray observations, the
limits which could arise from the corresponding cosmic ray
emission of the neutrino sources have been given little
attention. A detailed treatment of this problem regarding
predictions for neutrino fluxes from the decay of topological
defects (TD) has been given by Protheroe and Stanev
\cite{PS96prl}, and a brief discussion of the possible relevance
for diffuse neutrino fluxes from AGN by Mannheim
\cite{Man95app}.  Recently, it was proposed by Waxman \& Bahcall
\cite{WB99prd} that indeed the measured flux of ultra-high energy cosmic rays
provides the most restrictive limit on extragalactic diffuse
neutrino fluxes for a broad class of sources.  They claim that
this cosmic ray bound is for neutrinos of all energies two orders
of magnitude lower than the bound previously used which was based
on the EGRB.  Besides the obvious restriction of their result to
neutrinos from proton accelerators (i.e., excluding TD models),
their claim is mainly based on three assumptions: (i) neutrons
produced in photohadronic interactions can escape freely from
the source, (ii) magnetic fields in the Universe do not affect
the observed flux of extragalactic cosmic rays, and (iii) the
overall injection spectrum of extragalactic cosmic rays is
${\propto} E^{-2}$.  A key role is played by their assumption
(iii): By assuming a specific cosmic ray input spectrum, they can
normalize their bound at the ultra high energies, where they
argue that also (ii) applies.  Assumption (i) is justified by
showing that some particular sources of specific interest, like
the TeV-blazar Mrk\,501, or also GRB, are transparent to the
emission of neutrons.  The authors claim that this new bound set
by cosmic ray data essentially rules out the hypothesis that
hadronic processes in AGN jets can produce the EGRB, and
consequently that their neutrino fluxes are overestimated.

The purpose of the present paper is to revisit the role cosmic ray
observations can play to constrain models of neutrino production.  The paper
is organized as follows.  In Section\,\ref{sect:spectra}, we give a brief
review on the properties of photohadronic interactions, and derive the
production spectra for cosmic rays and neutrinos for power-law photon target
spectra.  In Section\,\ref{sect:prop} we briefly describe the effect of the
propagation of extragalactic cosmic rays.  We then follow Waxman \& Bahcall
in deriving a cosmic ray bound on neutrino fluxes, adopting their assumptions
(i) and (ii), but instead of assuming a specific cosmic ray injection
spectrum, we assume a specific spectrum for the {\em observable}
extragalactic cosmic ray flux, which is constructed such that it complies
with all existing observational limits on the cosmic ray proton intensity.
In Section\,\ref{sect:agn} we turn to AGN, and discuss in particular the
photohadronic opacity of blazar jets as can be estimated from observations.
We shall show that most AGN in fact have large photohadronic opacities at
ultra-high energies, and we derive an upper bound for the neutrino
contribution from AGN considering the effect of neutron opacity.  In
Section\,\ref{sect:magnetic}, we discuss the possible effect on cosmic ray
protons from neutron decay of the magnetic fields known to exist in clusters
of galaxies, and radio galaxies which are considered the hosts of gamma-ray
emitting AGN.  We derive critical energies below which they could increase
the bound.  We conclude by discussing the combined effect of our results, and
to what extent the cosmic ray data can indeed constrain models for expected
neutrino fluxes, and vice versa.

\section{Cosmic ray, gamma-ray, and neutrino emission from 
extragalactic proton accelerators}
\label{sect:spectra}

In this section, we obtain the form of the spectra of cosmic
rays, gamma-rays, and neutrinos escaping from cosmic proton
accelerators.  Here, we shall assume that protons are confined
within the acceleration region whereas photoproduced neutrons may
escape to become cosmic rays.  This is justified in particular
for cosmic ray sources connected to relativistic outflows, like
AGN jets or GRB, since the life time of protons is here limited
by adiabatic energy losses and generally is much shorter than the
diffusive escape time.  For example, if protons are accelerated near
the beginning of an AGN jet, say where the jet width is 
${\sim}10^{16}$~cm and where the gamma-rays are probably produced,
and are released near the end of the jet after it's width has
expanded to at least a few parsec,
then their energies on
release will be down by a least two orders of magnitude.  As a
consequence, only protons resulting from the decay of neutrons
which have escaped from the acceleration region contribute 
significantly to the
cosmic ray spectrum.  Escaping neutrons are produced by
accelerated protons which interact with ambient soft photons,
together with pions which decay into neutrinos.

We shall first discuss the properties of photohadronic
interactions.  We shall then obtain the ambient proton spectrum
in the emission region resulting from shock acceleration followed
by radiative cooling, adiabatic losses, and advection away from
the shocked region where the photon density is sufficient for
efficient photoproduction of neutrons.  From this we obtain the
form of the spectra of neutrons and neutrinos on production, and
the escaping neutron spectrum which may be modified by neutron
absorption in photohadronic $n\to p$ conversions.

\vspace{-1ex}
\subsection{Photohadronic interactions}
\label{phothad}
\vspace{-.5ex}

Photohadronic interactions can be divided into two processes:
photoproduction of pions (and other mesons), and Bethe-Heitler production of
$e^\pm$ pairs.  Charged pions decay as $\pi^\pm \to \mu^\pm\nu_\mu$,
$\mu^\pm\to e^\pm \nu_\mu\nu_e$ (here and in the following we disregard the
difference between neutrinos and anti-neutrinos), neutral pions decay into
gamma-rays as $\pi^0\to \gamma\gamma$.  Electrons and positrons (from pion
decay and Bethe-Heitler production) cascade in the magnetic field and
radiation field, and so can be assumed to convert all their energy into
synchrotron radiation in the magnetic field required for the acceleration of
protons.  The production of charged pions allows the production of secondary
neutrons through isospin exchange.

The physics of photohadronic interactions in ambient photon spectra has been
extensively studied in Monte-Carlo simulations
\cite{SOPHIA99pasa,SOPHIA99cpc}.  The properties of the production of
secondary particles can be expressed in the fraction $\xi$ of the proton
energy given to a specific particle component per interaction.  For
neutrinos, gamma-rays, and neutrons, the values
\begin{eqnarray}
\xi_\nu \approx \xi_\gamma &\approx& 0.1\\
\xi_n &\approx& 0.5 \;,
\end{eqnarray}
respectively, have been found for power law target spectra typical in AGN
jets, while for GRB target spectra the values are $\xi_\nu \approx \xi_\gamma
\approx \xi_n \approx 0.2$ \cite{SOPHIA99texas}.  The energy per particle in
units of the proton energy have been found for neutrinos and neutrons as
$\exval{E_\nu}/E_p \approx 0.033$ and $\exval{E_n}/E_p\approx 0.83$,
respectively, while for GRB they are $\exval{E_\nu}/E_p \sim 0.02$ and
$\exval{E_n}/E_p\sim 0.5$, respectively \cite{SOPHIA99texas}.  From this we can
immediately define the relative energy of escaping neutrinos and neutrons as
\begin{equation}
\eta_{\nu n} = \exval{E_\nu}/\exval{E_n} \approx 0.04 
\end{equation}
for both AGN and GRB target spectra.  The fractional energy loss of the
proton per interaction is $\kappa_p\approx 0.2$ in the AGN case, and
$\kappa_p\approx 0.5$ for GRB \cite{SOPHIA99texas}. Note that the quantities
given here as typical for GRB apply only at ultra-high proton energies, at
lower energies they approach the values found for AGN \cite{SOPHIA99texas}.

Electromagnetic radiation initiated by the Bethe-Heitler pair production, and
by photons and electrons from neutral and charged pion decay, are reprocessed
in synchrotron pair cascades.  This energy will emerge as a component of the
gamma ray background radiation, for AGN mainly in an energy range $10\MeV{-}
1\TeV$.  The contribution of the Bethe-Heitler process to the production of
gamma-rays depends on the target energy spectrum index, $\alpha$, since its
cross section peaks at energies about two orders of magnitude lower than that
of photopion production.  Assuming that the power law extends over this range
without change, one can find the relation
\begin{equation} 
\label{eq.gamma:nu}
L_\gamma = [1 + \exp(5\alpha - 5)] L_\nu\quad,   
\end{equation}
where $L_\gamma$ and $L_\nu$ are the bolometric photohadronic luminosities
in gamma rays and neutrinos \cite{RM98prd}, and we note that for $\alpha=1$,
$L_\gamma = 2 L_\nu$.  We also note that in general, $L_\gamma\ge L_\nu$
holds as a direct consequence of the isospin-symmetry of charged and neutral
pions --- hence, for any kind of neutrino production involving pion decay,
the bolometric flux in correlated photons sets an robust upper limit on the
possible bolometric neutrino flux.

\subsection{Ambient proton spectrum}

We assume a spectrum of protons on acceleration of the form, $Q(E_p) \propto
E_p^{-2} \exp(-E_p/E_{\rm max})$ [s$^{-1}$ GeV$^{-1}$].  
In order to
calculate the spectra of cosmic ray protons and neutrinos escaping from AGN
jets we first need to obtain the ambient spectrum of protons, $N_p(E_p)$
[GeV$^{-1}$], in the shocked regions where the soft photon target density is
sufficiently high for pion photoproduction to take place.  From the ambient
proton spectrum we can then obtain the spectra of neutrinos and neutrons
produced, $Q_n(E_n)$ and $Q_\nu(E_\nu)$ [s$^{-1}$ GeV$^{-1}$], respectively.
Then taking account of the optical depth of the emission region for neutrons
escaping photo-induced $n\to p$ conversions, one can obtain from the neutron
production spectrum, the spectrum of cosmic ray protons resulting from the
$\beta$-decay of the neutrons escaping from the jet, $Q_{\rm cr}(E_p)$.

The spectrum of protons on acceleration and the ambient proton spectrum are
related by the proton loss time scale, $t_p(E_p)$,
\begin{eqnarray}
N_p(E_p)
\sim Q_p(E_p) t_p(E_p)\;.
\end{eqnarray}
The processes mainly contributing to losses of protons are interactions with
radiation, advection away from the shock region of dimension $R$, and
adiabatic energy losses if the emission region expands.

The advection time scale is expected to be $t_{\rm adv} \sim R/(\beta_{\rm
sh} c)$ where $\beta_{\rm sh}$ is the shock velocity and $R$ is the dimension
of the jet in the shocked region.  In relativistic flows streaming away from
a central source, this energy independent time scale is usually in
competition with adiabatic energy losses of the protons due to the expansion
of the flow \cite{RM98prd}.  In a relativistic outflow, characterized by a
bulk Lorentz factor $\Gamma$ and an opening angle $\Theta$, the expansion
velocity in the co-moving frame (in units of $c$) is $\beta_{\rm ex} \approx
\Gamma\Theta$ for $\Theta<\Gamma^{-1}$, and $\beta_{\rm ex}\approx 1$
otherwise.  This leads to adiabatic cooling on a time scale $t_{\rm
ad}\approx R/(\beta_{\rm ex} c)$.  For example, in GRB one can generally
assume that $\beta_{\rm ex}\simeq 1$, and also observations of superluminal
motions in AGN jets are consistent with $\Theta \sim \Gamma^{-1}$
\cite{GD97apj}, thus $\beta_{\rm ex} \sim 1$.  In the following, we shall
assume that adiabatic losses are relevant with $0.3\lsim \beta_{\rm ex}\lsim
1$.  This has the important consequence that the lifetime of protons in the
jet (or outflow) is limited to about one crossing time.  The time scale for
diffusive escape of protons is usually much longer (except, maybe, near the
maximum proton energy), thus protons with $E_p\ll E_{\rm max}$ can be assumed
to be confined in the emitter and do not contribute to the cosmic ray
emission.

The photon target spectrum will be assumed to have a power law shape,
$n(\eps)\propto \eps^{-\alpha-1}$, extending to energies sufficiently above
the threshold for photopion production by protons of energy $E_p$.  Then, for
$\alpha>0$ the time scale for energy loss by photohadronic interactions is
asymptotically of the form
\begin{eqnarray}
\label{tpg}
t_{p \gamma}(E_p) \propto E_p^{-\alpha}\;,  
\end{eqnarray}
where $t_{p\gamma}$ is understood as including Bethe-Heitler and pion
production losses, the cooling time for pion production will be called
$t_{p,\pi} > t_{p\gamma}$.  For very flat target spectra, as for example in
GRB at ultra-high proton energies, the photoproduction time scale is
approximately constant \cite{RM98prd}, thus, Eq.~(\ref{tpg}) applies with
$\alpha=0$.  We shall confine the discussion to the values $\alpha=1$,
relevant in AGN, and $\alpha=0$ hereafter.  Hence, for $\alpha=1$ we obtain
\begin{eqnarray}
t_{p \gamma}(E_p) = (E_1/E_p)(R/c), 
\end{eqnarray}
with $t_{p,\pi}(E_p) \approx 1.5 t_{p\gamma}(E_p)$.  $E_1$ is the energy
corresponding to unit optical depth for photoproduction losses, $\tau_{p
\gamma}(E_1) = R/[c t_{p\gamma}(E_1)]=1$.   Setting $t_p^{-1} = t_{p, {\rm
ad}}^{-1} + t_{p \gamma}^{-1}$ we obtain,
\begin{eqnarray}
N_p(E_p)&=& Q_p(E_p) (R/c) [(E_p/E_1) + a]^{-1}
\end{eqnarray}
with $a = \max(\beta_{\rm ex},\beta_{\rm sh})$, and for $Q_p(E_p) \propto
E^{-2}$ we have
\begin{eqnarray}
N_p(E_p)& \propto & \left \{ \begin{array}{ll}
                a^{-1} E_p^{-2} & (E_p < a E_1)\\
                E_1 E_p^{-3} & (E_p > a E_1)
                \end{array} \right.  .
\end{eqnarray}
(Note that for clarity, here and in the next section we omit the exponential
cut-off in the spectrum at $E_{\rm max}$.) Obviously, for $\alpha=0$ the
optical depth for photoproduction is constant, and $N_p(E_p) \propto
Q_p(E_p)$.  One can show that for typical photon densities in GRB fireballs
$\tau_{p\gamma} < 1$, and that Bethe-Heitler losses are unimportant, viz.,
$t_{p,\pi} \approx t_{p\gamma}$ \cite{WB97prl,RM98prd}.

\subsection{Generic cosmic ray proton and neutrino production spectra}
\label{CGN_generic}

The time scale for photohadronic production of neutrons is $t_{p \gamma \to
n} \approx t_{p,\pi}\kappa_p/\exval{N_n}$.  For $\alpha=1$, this is
$t_{p\gamma\to n} \approx 0.5 t_{p \gamma} \propto E_p^{-1}$, while for
$\alpha =0$ we have $t_{p\gamma\to n} \approx 2.5 t_{p \gamma}$.  This
immediately gives the production spectrum of neutrons,
\begin{eqnarray}
Q_n(E_n) & \approx &  N_p(E_n)/t_{p \gamma \to n}(E_n)\nonumber\\
& \propto & \left \{ \begin{array}{ll}
                a^{-1} E_n^{-1} & (E_n < aE_1)\\
                E_1 E_n^{-2} & (E_n > aE_1)
                \end{array} \right.  .
\end{eqnarray}
Neutrons may escape to become cosmic ray protons.  However, because neutrons
themselves suffer pion photoproduction losses, the cosmic ray production
spectrum will differ from $Q_n(E_p)$ above the energy, $b E_1$ at which the
optical depth for neutron escape, $\tau_{n\gamma}$, is one.  Neutrons can be
considered as ``absorbed'' after they are converted into a proton, or after
they have lost most of their energy in $n\gamma$ interactions, whichever time
scale is shorter.  For $\alpha=1$ this means $\tau_{n\gamma}\approx 2
\tau_{p\gamma}$, giving $b \approx 0.5$ for AGN jets, while for $\alpha=0$
and typical GRB photon densities, $\tau_{n\gamma}\approx \tau_{p\gamma} < 1$,
which means that neutron absorption is unimportant in GRB.

We note that in a homogeneous spherical medium of radius $R$ the
optical depth decreases radially $\propto(1-r/R)$ from its central 
value $\tau$ to zero at $r=R$ giving rise
to the geometrical escape probability of an interacting particle propagating
in a straight line
\begin{eqnarray}
\cP_{\rm esc}(\tau) & \approx & (1-e^{-\tau})/\tau \nonumber\\
& \approx & \left \{ \begin{array}{ll}
                1       & \tau < 1 \\
                \tau^{-1}   & \tau > 1 
                \end{array} \right.  ,
\label{eq.pesc}
\end{eqnarray}
resulting in the cosmic ray proton production spectrum being steepened above
$bE_1$, compared with $Q_n(E_p)$.   For $\alpha=1$, the cosmic ray proton
production spectrum is therefore
\begin{eqnarray}
Q_{\rm cr}(E_p) 
& \propto & \left \{ \begin{array}{ll}
                a^{-1} E_p^{-1}E_1^{-1} & (E_p < aE_1)\\
                E_p^{-2} & (aE_1 < E_p < bE_1)\\
                bE_1 E_p^{-3} & (bE_1 < E_p)
                \end{array} \right.  .
\end{eqnarray}
However, because the two break energies, $aE_1$ and $bE_1$, are 
probably very close, $a \sim b$, we shall adopt a single break energy
$E_b=bE_1$, and use the following approximations
\begin{equation}
Q_n(E_n,L_p) \propto L_p \exp\left[\frac{-E_n}{E_{\rm max}}\right]
		\left \{ \begin{array}{ll}
                E_n^{-1}E_b^{-1} & (E_n < E_b)\\
                E_n^{-2} & (E_b < E_n)
                \end{array} \right.  ,
\end{equation}
\begin{equation}
Q_{\rm cr}(E_p,L_p) \propto L_p \exp\left[\frac{-E_p}{E_{\rm max}}\right]
		\left \{ \begin{array}{ll}
                E_p^{-1}E_b^{-1} & (E_p < E_b)\\
                E_p^{-3}E_b      & (E_b < E_p)
                \end{array} \right.  .
\label{eq.generic_cr}
\end{equation}
Note that we have now put in explicitly the cut-off in the accelerated proton
spectrum, and the proportionality with the proton luminosity of the source,
$L_p$.  In Sect.\,\ref{sect:agn}, we shall relate $L_p$ to the to observed
photon luminosities for specific models of neutrino emission by AGN jets.
Obviously, in GRB we simply have $Q_{\rm cr}(E_n,L_p) \approx Q_n(E_n,L_p)
\propto L_p E_p^{-2} \exp(-E_p/E_{\rm max})$.

The production spectrum of muon-neutrinos will have the same broken power-law
form as the neutron production spectrum, and is related to it by
\begin{equation}
Q_{\nu_\mu}(E) 
        = \frac23 \frac{\exval{\xi_\nu}}{\exval{\xi_n}\eta^2_{\nu n}} 
                Q_n(E/\eta_{\nu n}) 
\label{eq.generic_nu}
\end{equation}
where we count $\nu_\mu$ and $\bar\nu_\mu$ together, and the
corresponding spectrum of electron neutrinos at the source would
be $Q_{\nu_e}(E) \approx \frac12 Q_{\nu_\mu}(E)$.  Putting in the numbers
given in Sect.\,\ref{phothad}, we find 
\begin{eqnarray}
\label{generic:AGN} 
        Q_{\nu_\mu}(E) \approx& 83.3 Q_n(25 E) &\for \alpha = 1\\
\label{generic:GRB} 
        Q_{\nu_\mu}(E) \approx& 416.  Q_n(25 E) &\for \alpha = 0\quad.   
\end{eqnarray}
We shall refer to Eq.~(\ref{eq.generic_cr}) and Eq.~(\ref{eq.generic_nu}) as the
{\em generic cosmic ray and neutrino production spectra}.  We emphasize the
strong dependence of the number of produced neutrinos per produced neutron on
the assumed target photon spectral index: at ultra-high energies, GRB produce
about 5 times more neutrinos per neutron than AGN.  We shall return to the
implications of this result at the end of the paper.

\section{Propagation of neutrinos, photons, and protons over 
cosmological distances}
\label{sect:prop}

In this section, we discuss propagation of cosmic rays and neutrinos in an
expanding Universe filled with the cosmic microwave background radiation.   To
illustrate the problem, we compare of energy-loss horizons of protons and
neutrinos.  We shall then briefly discuss the physical problems connected to
several approaches to cosmic ray propagation calculations.  Using the
numerical propagation code described by Protheroe and Johnson \cite{PJ96app},
we then calculate the observable neutrino and cosmic ray spectra from a
cosmological distribution of generic photohadronic sources, as described in
the last section.  Here we assume that the sources are transparent to
neutrons, while protons are confined, and that gamma-rays are reprocessed in
synchrotron-pair cascades until emitted in the energy range of
$10\MeV{-}30\GeV$.  Using an extrapolated cosmic ray spectrum which is
consistent with present observational limits on the light component of
cosmic rays (i.e.\,protons) as an upper limit on the possible extragalactic
proton contribution, and the diffuse extragalactic gamma ray background
(EGRB) observed by EGRET as an upper limit on the hadronic extragalactic
gamma ray flux, we determine an energy dependent upper bound on the neutrino
flux from cosmic ray sources with the assumed properties.  Our result is
compared with the energy independent bound on extragalactic neutrino fluxes
recently proposed by Waxman\,\&\,Bahcall\,\cite{WB99prd}.

\subsection{Comparison of energy-loss horizons}

We wish to compare the distances that neutrinos, photons with energies below
threshold for cascading in background radiation fields, and protons will
travel through the Universe without significant energy losses.   We define the
energy loss horizon by
\begin{equation}
\lambda = c\,E/|dE/dt|,
\end{equation}
such that for linear processes the energy is reduced to $1/e$ of its initial
value on traversing a distance $\lambda$.

For gamma-rays below ${\sim}30\GeV$ and neutrinos, the energy-loss process is
due to expansion of the Universe \cite{MP96apj}.  For simplicity, we adopt an
Einstein-de\,Sitter cosmology (i.e., $\Lambda=0$ and $\Omega=1$), so that the
horizon $\lambda$ can be related to a redshift $z$ by the redshift-distance
relation
\begin{equation}
\label{eq.lambda:z}
\lambda(z) = \frac23 \frac{c}{H_\circ} \left(1-(1+z)^{-3/2}\right)
\end{equation}
where $H_\circ = 50 h_{50} \km\scnd^{-1}\Mpc^{-1}$ is the Hubble
constant.   Since we require the distance for which the energy
is reduced by a factor $e$ during propagation, we have $(1+z)=e$
giving the horizon for redshift losses,
\begin{equation}
\lambda_z = \frac{2c}{3 H_\circ}(1 - e^{-3/2}) .  
\end{equation}
This is also the horizon for neutrinos and gamma-rays below ${\sim} 30\GeV$.
Normalizing the neutrino horizon to the radius of the
Einstein-de\,Sitter universe,
\begin{equation}
\hat\lambda_z \equiv \frac{3 H_\circ}{2c} \lambda_z,
\end{equation}
we obtain $\hat\lambda_\gamma = \hat\lambda_\nu = \hat\lambda_z \approx
0.78$.

In addition to redshift losses, extragalactic cosmic rays suffer energy
losses from photohadronic interactions with cosmic backgrounds, mainly the
microwave background, and this is the reason for the Greisen-Zatsepin-Kuzmin
(GZK) cut-off expected for a cosmic ray spectrum originating from a
cosmologically homogeneous source distribution \cite{Gre66prl,ZK66jetpl}.   
Photo-pion production and Bethe-Heitler pair production govern the energy
loss in different energy regimes due to their very different threshold
energies.   The Bethe-Heitler process limits the propagation of protons with
energies $E_p > 2 m_p m_e c^4/k T_{\rm mbr} \approx 4\mal 10^9\GeV$ to
$\lambda_p \sim 1\Gpc$, while pion production reduces the horizon for protons
with $E_p \gsim m_p m_\pi c^4/k T_{\rm mbr} \approx 5\mal 10^{11}\GeV$ to
$\lambda_p\sim 10\Mpc$.   Again, in units of the radius of the Einstein-de
Sitter Universe, the energy-dependent horizon for protons can be written
\begin{equation}
\hat\lambda_p(E) = [1/\hat\lambda_z + 1/\hat\lambda_{p,\rm BH}(E) +
1/\hat\lambda_{p,\pi}(E)]^{-1}\;,
\end{equation}
where the components expressing redshift, Bethe-Heitler and pion
production losses, can be written as
\begin{eqnarray}
\hat\lambda_z     &\simeq& 0.78 \nonumber\\   
\label{eq.lambdap:approx}
\hat\lambda_{p,\rm BH}(E) &\approx& 0.27 h_{50} \exp(0.31/E_{10}) \\
\hat\lambda_{p,\pi}(E)    &\approx& 5\mal 10^{-4} h_{50} \exp(26.7/E_{10})
\quad,\nonumber
\end{eqnarray}
with $E_{10} = E_p/10^{10}\GeV$.  The approximations for $\hat\lambda_{p,\rm
BH}$ and $\hat\lambda_{p,\pi}$ fit the exact functions determined numerically
in \cite{PJ96app} and the exact interaction kinematics within $\sim10\%$ up to
$E_p\sim 10^{12}\GeV$.

The different energy-loss horizons for gamma rays and neutrinos, and protons
strongly affect the relative intensities of their diffuse isotropic
background fluxes.  This is true in particular for evolving source
populations such as quasars, galaxies, or GRBs (if they trace star formation
activity), since here most of the energy is released at large redshifts.
Cosmic rays above the ankle ($E_{\rm cr}\sim 3\mal 10^9\GeV$) originate only
from sources with redshifts $z \lsim 0.27$, while neutrinos and gamma rays
originate from sources within $z_\nu = z_\gamma \sim 1.7$.  This will give
rise to the neutrino intensity being enhanced relative to the protons because
of their larger horizon, and because of the evolution of the sources (e.g.
quasars) with cosmic time (redshift).

We may illustrate the problem as follows.  The basic method of calculating the
approximate present-day diffuse fluxes of neutrinos, gamma rays below ${\sim}
30\GeV$, and cosmic rays of photoproduction origin, would be to integrate the
contributions from sources at redshifts up to those corresponding to the
respective energy-loss horizons.  Assuming a constant source number per
co-moving volume element for simplicity, the resulting fluxes are proportional
to $V_{\rm c}(\hat\lambda)/d_{\rm L}^2(\hat\lambda) \sim \hat\lambda$, where
$V_c(\hat\lambda)$ and $d_L(\hat\lambda)$ are the cosmological co-moving
volume and the luminosity distance, respectively, corresponding to the
horizon $\hat\lambda$.  Assuming photohadronic production of neutrinos and
cosmic rays in, e.g., a cosmological (non-evolving) distribution of AGN, the
relative flux of neutrons (assuming no absorption) with energy $E_n$, and
corresponding neutrinos with energy $E_\nu = 0.04 E_n$, is at the source
given by $[E_\nu^2 N_\nu(E_\nu)]/[E_n^2 N_n(E_n)] = \xi_\nu/\xi_n \approx
0.2$.  The same ratio will be observed in the integrated fluxes, as long $E_n
< 4\mal10^9\GeV$.  For higher cosmic ray energies, the flux ratio must be
multiplied by a factor $\hat\lambda_\nu/\hat\lambda_p$, which yields a flux
ratio of ${\sim} 0.6$ for $E_n\sim 10^{10}\GeV$, and ${\sim} 30$ for $E_n\sim
10^{11}\GeV$.  Obviously, the differences would be much larger if we had
assumed strong source evolution which enhances the contribution from large
distances.  If we want to determine a neutrino spectrum from an observed,
correlated cosmic ray spectrum from the same sources, the result must
therefore approximately reflect the changes in the ratio of the energy loss
horizons.

\subsection{Exact calculation of present-day neutrino and cosmic ray spectra}

The example above, of simply integrating the cosmic ray and neutrino
contribution of cosmologically distributed sources up to the energy-loss
horizon, disregards several important aspects of particle propagation.
Firstly, the particle number is not conserved in this method.  Secondly, the
energy evolution of the particles is neglected, which removes the dependence
on spectral properties of the source.  Both caveats are removed in a method
known as the continuous-loss approximation, which follows the particle energy
along fixed trajectories as a function of cosmological distance (or redshift)
\cite{BG88aa,RB93aa}.  For example, the trajectories for neutrino energies
would be simply $E(z) = E_0(1+z)$.  This method is exact for adiabatic losses
due to the expansion of the Universe (i.e., particle redshift), and is still
a very good approximation for Bethe-Heitler losses, but it gives only poor
results for photo-pion losses.  The reason for the latter is the large mean
free path, and the large inelasticity of this process, which results in
strong fluctuations of the particle energy around its mean trajectory
\cite{HS85prd,Rac96PhD}.  Cosmic ray transport in the regime where photo-pion
losses are relevant is therefore best described by numerical approaches,
either solving the exact transport equation \cite{HS85prd}, or by Monte-Carlo
simulations \cite{YT93ptp}.  Another important aspect concerning relative fluxes
of cosmic rays, gamma rays and neutrinos is the fact that the interaction of
the cosmic rays with the cosmic background radiation themselves produces
secondary particles, which have to be considered as an additional
contribution to the primary neutrinos and gamma rays.  The full problem is
treated by the cascade propagation code which has been described in detail by
Protheroe and Johnson \cite{PJ96app}, and which we shall use also here.  The
code is based on the matrix-doubling technique for cascade propagation
developed by Protheroe and Stanev \cite{PS93mnras}.

For an input spectrum $(dP_{\rm gal} / dV_c)\langle Q(E,z)\rangle$ per unit
co-moving volume per unit energy per unit time, the intensity at Earth at
energy $E$ is given by
\ifonecol\begin{equation}\else\begin{eqnarray}\fi
\label{CRint}
I(E) \ifonecol \propto\frac1{4\pi} \else &\propto&\\&\DS\frac1{4\pi}& \fi
	\int\limits_{z_{\rm min}}^{z_{\rm max}} \!\!\! M(E,z) 
	{(1+z)^2 \over 4 \pi d_L^2} {{\rm d} V_c \over {\rm d}z} 
	\frac{{\rm d}P_{\rm gal}}{{\rm d}V_c}
	\big\langle Q[(1+z)E,z]\big\rangle \, {\rm d}z
\ifonecol\relax\else\nonumber\fi
\ifonecol\end{equation}\else\end{eqnarray}\fi
where $d_L$ and $V_c$ are luminosity distance and co-moving volume, and
$M(E,z)$ are ``modification factors'' for injection of protons at redshift
$z$ as defined by Rachen and Biermann \cite{RB93aa}; for neutrinos,
$M(E,z)=1$.  The modification factors for protons depend on the input spectra,
and are calculated numerically using the matrix method \cite{PJ96app}.
%
%
\def\captionfigone{The observed all-particle cosmic ray spectrum spectrum
taken from the article by T.K.~Gaisser and T.~Stanev in the 1998 Review of
Particle Properties \protect\cite{PDG98}, and supplemented by Fly's Eye
monocular data \protect\cite{BCD+94apj} (open circles at high energy), and
recent AGASA data \protect\cite{THH+98prl} (filled circles at high energy).
Also shown are estimates of the cosmic ray proton component: based on the
proton fraction estimates by \protect\cite{GST+93comap} (thick lines with
thick error bars, extended by thin lines which indicate the systematic error
due to normalization to the all-particle data); Norikura data
\protect\cite{SYK+93app} (filled circles with large error bars at $3 \times
10^6-10^7\GeV$); proton fraction estimated from KASCADE data
\protect\cite{KASCADE99proc} normalized to all-particle data (hatched band
from $10^6-10^7\GeV$).  The spectrum we adopt for the proton component which
forms the upper bound to any extragalactic cosmic ray proton spectrum
[cf.~Eq.~(\protect\ref{crmax})] is shown by the dotted line.}
\ifsubmit\else
\begin{figure}[t]
\centerline{\kern0.4cm\epsfysize7.1cm\epsfbox{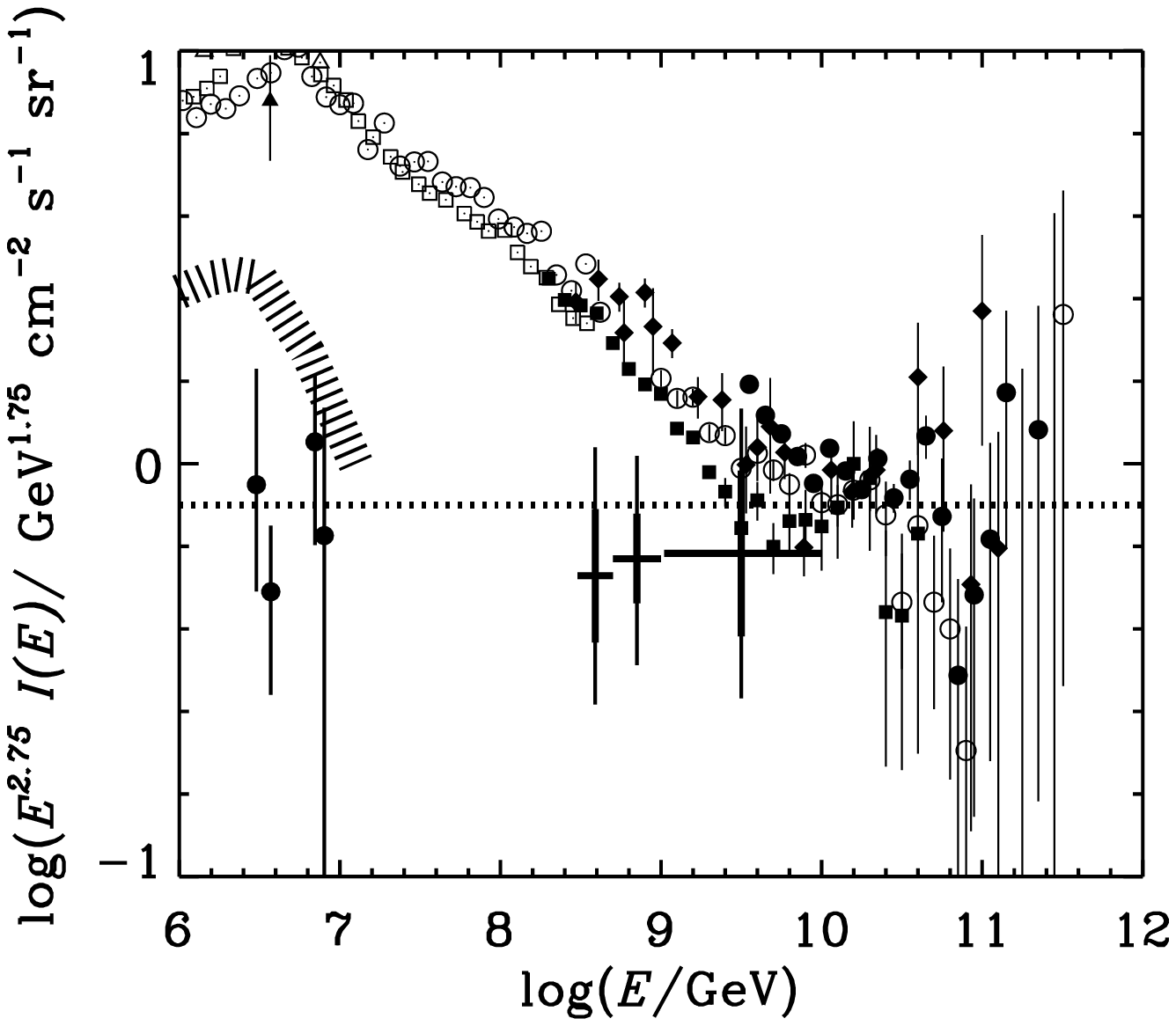}}
\caption[]{\label{fig1}\capsize\captionfigone}
\end{figure}
\fi 
%

\subsection{An abstract bound on extragalactic neutrino fluxes from
neutron-trans\-parent sources}
\label{sect:thinbound}

From the above considerations, it is obvious that one can use the observed
cosmic ray spectrum to construct a correlated neutrino spectrum, under the
assumption that all observed extragalactic cosmic rays are due to neutrons
ejected from the same sources as the neutrinos.  A difficulty here is that
the contribution of {\em extragalactic} cosmic rays to the total observed
cosmic ray flux is unknown.  Since neutrons convert into cosmic ray protons,
we can clearly consider only the proton component at all energies.  Above the
ankle in the cosmic ray spectrum at $E_{\rm cr}\sim 3\mal10^9\GeV$,
observations are generally consistent with a ``light'' chemical composition,
i.e., a $100\%$ proton composition is possible.  Since protons at these
energies cannot be confined in the magnetic field of the Galaxy, they are
also likely to be of extragalactic origin.  At extremely high energies,
however, there is the problem that the event statistics is very low, and
different experiments disagree on the mean cosmic ray flux at 
${\sim}10^{11}\GeV$ by one order of magnitude (see Bird et al.\ \cite{BCD+94apj} for
a comparison of the results until 1994 of the four major experiments Akeno,
Havarah Park, Fly's Eye, and Yakutsk). This energy region is very important,
since we expect here the existence of the GZK cutoff due to photoproduction
losses in the microwave background.  Currently, no clear evidence for the
existence of this cutoff has been found, and the results of at least two
major experiments (Havarah Park \cite{LRW91jpg}, and AGASA 1998
\cite{THH+98prl}) are consistent with the result obtained from a
superposition of all experiments using a maximum likelihood technique
\cite{Sigl95sci,Wat98npbps}, that is, a continuation of the cosmic ray
spectrum as a power law ${\propto} E^{2.75}$ up to ${\gsim}3\mal 10^{11}\GeV$
(see also Fig.~\ref{fig1}).
%
%
\def\captionfigtwo{Spectra of (a) cosmic rays and (b) neutrinos
after propagation through the Universe of input spectra for optically thin
pion photoproduction sources for $E_{\rm max}=10^6, 3\times 10^7, 10^9,
3\times10^{10}, 10^{12}$, and $3 \mal 10^{13}\GeV$, assuming galaxy evolution
as described in the text.  Spectra are normalized such that the cosmic ray
intensity does not exceed the cosmic ray proton spectrum estimated from
observations (dotted line in part a) and such that the neutrino energy flux
does not exceed 0.5 of the observed photon energy density above $3\MeV$.  The
dotted curve in part b joins the peaks in the neutrino spectra and forms our
neutrino upper bound for optically thin pion photoproduction sources.  The
dashed line is the bound obtained by Waxman and Bahcall 
\protect\cite{WB99prd}.}
\ifsubmit\else
\ifrevtex\begin{figure*}[t]\else\begin{figure}[tp]\fi 
\centerline{\kern0.4cm\epsfysize7.1cm\epsfbox{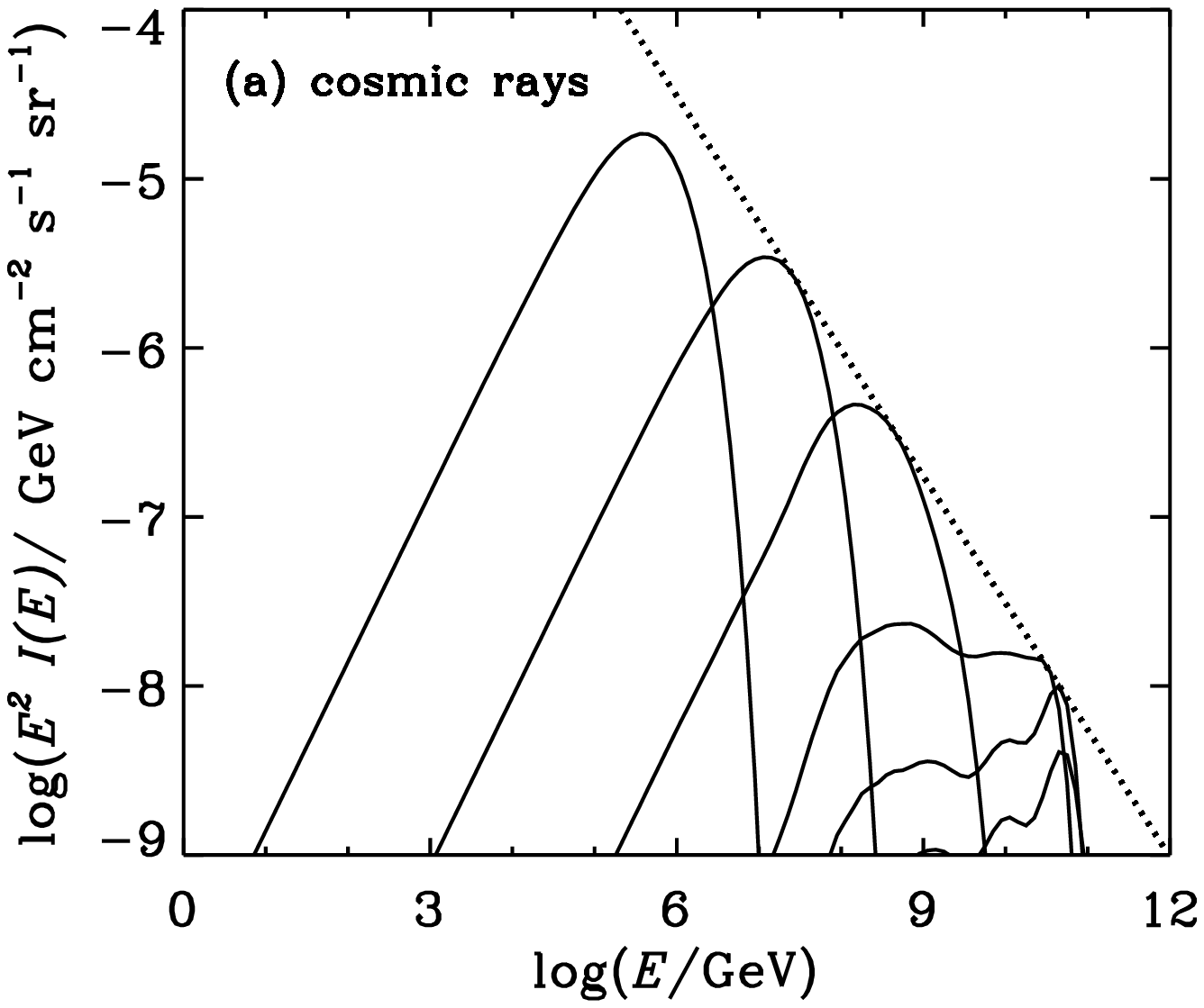}\hss
        \epsfysize7.1cm\epsfbox{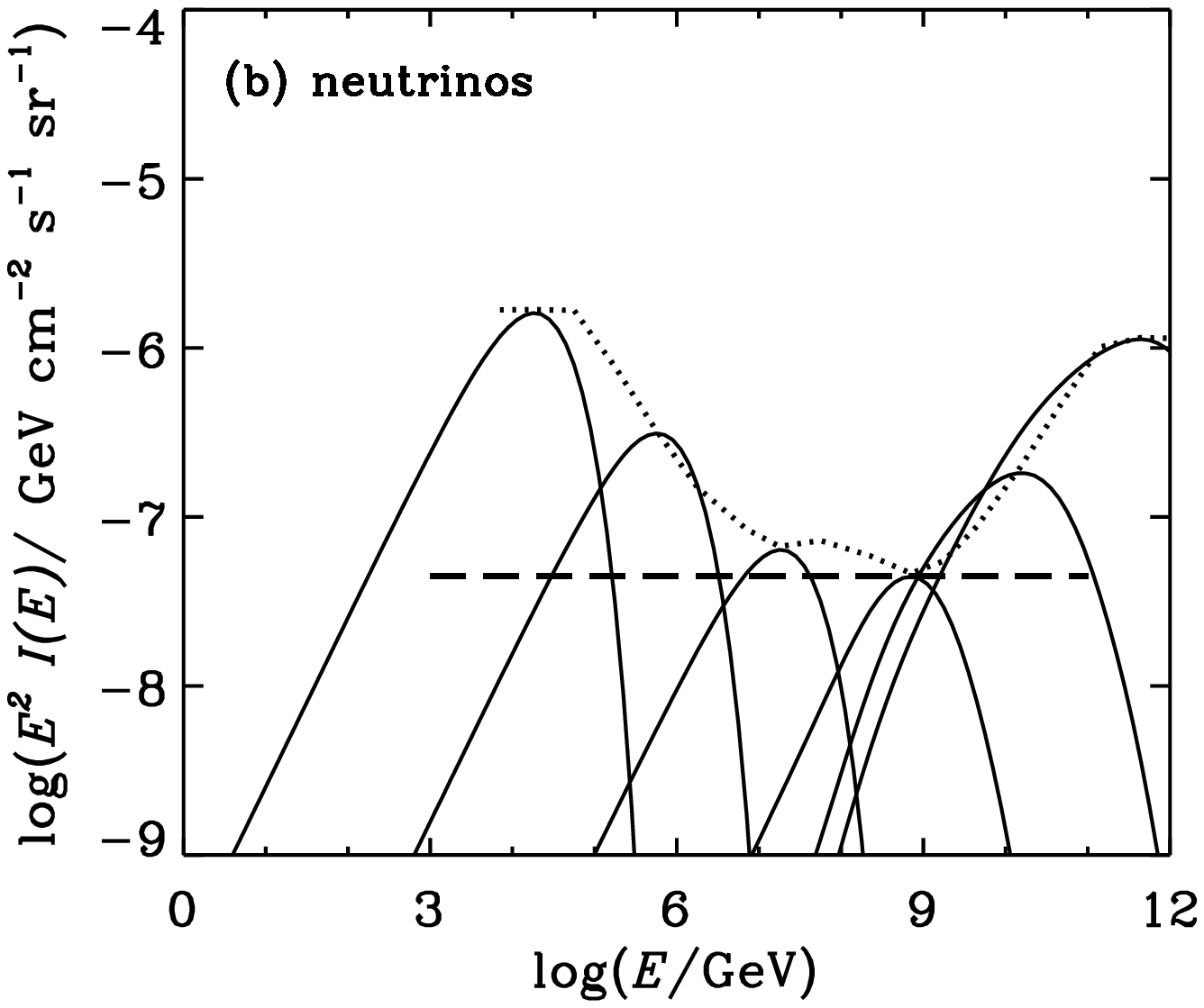}} 
\caption[]{\label{fig2}\capsize\captionfigtwo}
\ifrevtex\end{figure*}\else\end{figure}\fi 
\fi 
%

The situation is even more difficult at lower energies: cosmic rays are here
assumed to be mainly of Galactic origin, and there is evidence that a
considerable, maybe dominant fraction consists of heavy nuclei rather than
protons.  Around the knee or the cosmic ray spectrum at $E_{\rm cr}\sim
10^{6}$--$10^7$ GeV, recent results from the KASCADE air shower experiment
suggest that the fraction of heavy nuclei in the cosmic ray flux is at least
${\sim} 30\%$, and further increasing with energy \cite{KASCADE99proc}.  Also
below the ankle, in the energy range $10^{8}{-}10^{9}$ GeV, the analysis of
air shower data has produced tentative evidence of a composition change from
heavy to light (with increasing energy), supporting a dominantly heavy
composition of cosmic rays between the knee and the ankle
\cite{GST+93comap}.  (Note that this result is under dispute, and it has
been shown that it depends on the Monte-Carlo simulation codes used to
construct the air-shower properties in dependence of the primary particle
mass \cite{DMS98app}.  These simulation codes involve particle interaction
models based on extrapolations many orders of magnitude above the energy
range currently accessible with particle accelerators \cite{KASCADE99jpg}.)

Using all the available data, we find that an extragalactic cosmic ray
spectrum of the form 
\begin{eqnarray}
\label{crmax}
N_{p,\rm obs}(E) &=& 0.8 \times (E/{\rm 1 \, GeV})^{-2.75}\; {\rm cm^{-2} \,
s^{-1} \, sr^{-1} \, GeV^{-1}} \nonumber\\ 
& & \quad (3 \times 10^{6}\GeV <E< 10^{12}\GeV)
\end{eqnarray}
is consistent with all data and limits on the cosmic ray proton flux
(Fig.~\ref{fig1}).  It represents the current {\em experimental upper limit}
on the extragalactic cosmic ray proton flux, which we shall use to construct
an {\em upper limit} on the possible, diffuse extragalactic neutrino flux.
If it can be shown that the intensity of protons at $10^{9}\GeV$ and lower
energies is below that assumed in this paper (dotted line in Fig.~1), then
the neutrino bound we have constructed below $10^7\GeV$ would need to be
reduced.

To construct the neutrino bound, we assume test spectra of the
form $Q_{\rm cr}(E) = Q_n(E) \propto E^{-1}\exp(-E/E_{\rm max})$
with $10^6\GeV{<} E_{\rm max}{<}10^{12}\GeV$.  The corresponding
neutrino spectra are determined using Eq.~(\ref{generic:AGN}).
We assume a source distribution following the cosmological
evolution function found for galaxies and AGN (\cite{BT98mnras},
see next section).  For a given $E_{\rm max}$, the total
contribution of cosmic rays and neutrinos is calculated using
Eq.~(\ref{CRint}).  The resulting spectrum is then normalized so
that its maximum reaches the cosmic ray flux given by
Eq.~(\ref{crmax}).  By varying $E_{\max}$ between $10^6\GeV$ and
$10^{12}\GeV$, we then obtain the desired maximum flux of
neutrinos consistent with the present cosmic ray data (see
Fig.~\ref{fig2}).  We also consider the correlated gamma ray
output, assumed to be twice the neutrino energy flux, and check
it does not exceed the observed power-law component of the
diffuse gamma-ray background (we estimate the background between
3 MeV and 30 GeV to be ${\approx}1.5 \mal 10^{-5}$ GeV cm$^{-2}$
s$^{-1}$ sr$^{-1}$, \cite{SBD+98apj}).  Note that the cosmic ray
curve for $E_{\rm max}=10^6\GeV$ does not reach the estimated
cosmic ray proton spectrum in Fig.~\ref{fig2}(a) in order to
avoid over-producing diffuse gamma rays.  Note also that the
propagated cosmic ray spectra cut off at or below $10^{11}\GeV$
for all $E_{\rm max}$ values, and so our bound is insensitive to
the assumed extragalactic cosmic ray spectrum above $10^{11}\GeV$.

The upper bound on the neutrino flux according to this model is
given by the minimum of the cosmic ray and the gamma ray bound,
shown in Figure \ref{fig3} by the curve ``$\tau_{n\gamma}<1$''.
We see that the cosmic ray limit starts to dominate the bound for
$E_\nu \gsim 100\TeV$, which then decreases to a minimum at
$E_\nu \approx 10^9\GeV$, after which it rises again.  This rise
is a consequence of the strongly increasing ratio of neutrino and
proton horizons above this energy, in conjunction with the
assumptions that the distribution of sources is homogeneous in
space.  A more realistic scenario might be that the lack of
evidence for the GZK cutoff in the cosmic ray data is due to the
dominant contribution of a local cosmic ray source.  In this
case, our estimate that $[E_\nu^2 N_\nu(E_\nu)]/[E_n^2 N_n(E_n)]
\propto
\hat\lambda_\nu/\hat\lambda_p$ would not apply.  Obviously, if an extended
data set would confirm the existence of the GZK cutoff in the UHECR spectrum,
the neutrino bound would remain at the level of the Waxman-Bahcall result
(also shown in Fig.~\ref{fig3}, see discussion below) for $E_\nu>10^9\GeV$.
%
%
\def\captionfigthree{Muon neutrino upper bounds for optically thin pion
photoproduction sources (curve labeled $\tau_{n\gamma}<1$) and optically
thick pion photoproduction sources (curve labeled $\tau_{n\gamma}\gg1$); the
hatched range between the two curves can be considered the allowed region for
upper bounds for sources with $\tau_{n\gamma}>1$.  For comparison we show the
bound obtained by Waxman and Bahcall \protect\cite{WB99prd} (for an evolving
source distribution).  Predictions for optically thin photoproduction sources
are also shown: proton-blazar (Mannheim 1995 \protect\cite{Man95app}, Model
A) -- dotted curve, and GRB sources (Waxman and Bahcall 1997,
\protect\cite{WB97prl}) -- dashed curve.  Also shown is an observational
upper limit from Fr\'ejus \protect\cite{Frejus96app} and the atmospheric
background \protect\cite{Lip93app}.}
\ifsubmit\else
\begin{figure}[tp]
\centerline{\kern0.4cm\epsfysize7.1cm\epsfbox{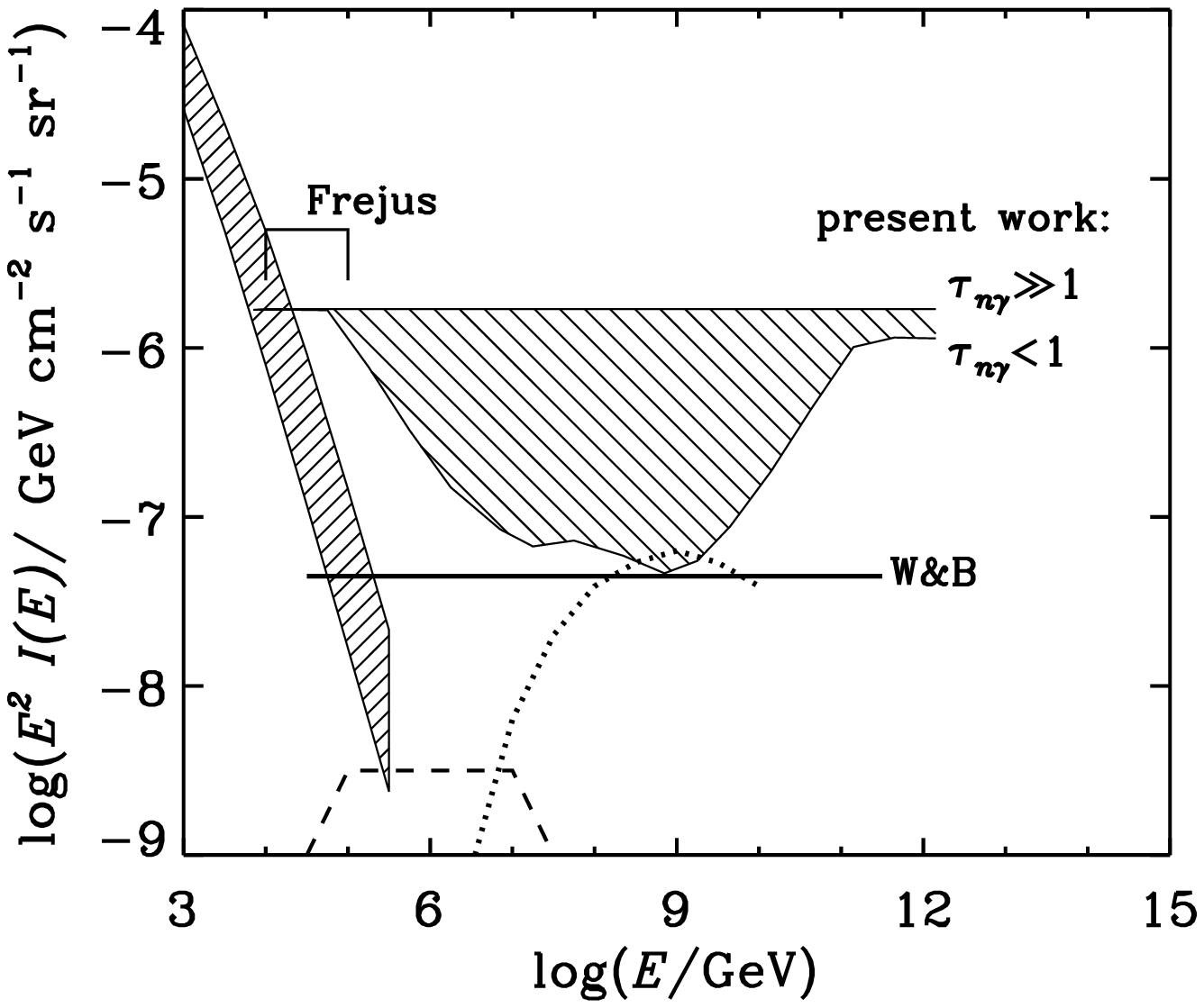}}
\caption[]{\label{fig3}\capsize\captionfigthree}
\end{figure}
\fi 
%

By assumption, the neutrino bound constructed this way applies only to
sources which are transparent to neutrons.  For the opposite extreme, i.e.,
sources with a very high neutron opacity, $\tau_{n\gamma}\gg 1$, it is still
possible to set an upper limit using the observed EGRB, assuming that the
dominant part of the emitted gamma-radiation is in the EGRET range.  This is
shown by the line labeled ``$\tau_{n\gamma}\gg 1$'' in Figure \ref{fig3}.
The range in between can be regarded the ``allowed range'' for the neutrino
emission from sources with $\tau_{n\gamma}>1$.  In the next section, we will
estimate specific neutrino spectra for AGN models which imply that
$\tau_{n\gamma}(E_n)>1$ for high neutron energies.

Our result may be compared with the cosmic ray bound to extragalactic
neutrino fluxes recently proposed by Waxman \& Bahcall \cite{WB99prd}. 
Their bound was constructed using a different approach: Waxman \&
Bahcall assume an $E^{-2}$ input spectrum of extragalactic
cosmic rays, and normalize the propagated spectrum to the observed flux
at $E_{\rm cr} = 10^{10}\GeV$.  Consequently, their bound (determined
for a source evolution $\propto (1+z)^{3.5}$) agrees with the one
derived in this work at $E_\nu\sim 5\mal 10^8\GeV$, where the cosmic ray
limit is most  restrictive.  

We have chosen $E^{-1}$ trial spectra with variable exponential cutoffs in
order to be able to mimic the effect of the superposition of spectra from
various source classes. The pronounced peak in the energy flux associated
with the trial spectra allows one to normalize the neutrino flux consistent
with the experimental upper limit on extragalactic protons ($\propto
E^{-2.75}$) at any chosen energy. Other hard spectra or delta-function
distributions as trial spectra would have yielded practically the same
result.  Although canonical AGN jets, which we discuss in the next section,
are an example for sources with $E^{-1}$ (or similar) spectra, our result
does not imply that we assume AGN jets to actually saturate the upper limit
in general. As a matter of fact, at neutrino energies ${\gsim}10^9\GeV$, a
class of photohadronic sources saturating our bound has neither been
suggested nor does its existence seem likely on the basis of current
knowledge. By restricting their source spectra to $E^{-2}$,
Waxman\,\&\,Bahcall have constructed a bound for neutron-transparent sources
which is probably closer to current models for cosmic ray and neutrino
production than our general upper limit above $10^{19}$~eV.  Examples for
such models are the model for diffuse neutrino fluxes from GRBs proposed by
the same authors \cite{WB97prl,WB99prd} (note that this prediction assumes no
cosmological evolution for GRBs), and Model A in Mannheim~\cite{Man95app}
which was constructed using the cosmic ray limit with the assumption that the
emerging neutrons at ${\sim} 10^{10}\GeV$ contribute to the extragalactic
cosmic ray spectrum (both shown in Fig.~\ref{fig3}).

\vspace{-1ex}
\section{Diffuse neutrino spectra from AGN jets} 
\label{sect:agn}
\vspace{-.5ex}

In this section, we shall consider the spectra of cosmic ray protons and
neutrinos emerging from jets of two classes of gamma-ray emitting AGN, i.e.,
BL Lac objects and radio-quasars, which are usually combined as the class of
{\em blazars}.  We shall use the estimated optical depths of gamma rays to
photon-photon pair production in a typical AGN of each type, to infer the
corresponding neutron-photon optical depths in these objects.  We shall show
that high luminosity AGN (like 3C\,279) can be expected to be opaque to
neutrons at energies above about $10^8{-}10^9\GeV$, while low luminosity BL
Lacs (like Mrk\,501) must be transparent to neutrons at all energies.  Then
assuming a model for the luminosity dependence of the optical depths, and the
local luminosity functions of BL Lacs and quasars, we shall estimate the form
of the production spectra of cosmic ray protons and neutrinos per unit volume
of the local Universe.  Applying the source evolution functions found for BL
Lacs and radio-quasars, respectively, we shall derive model estimates for the
diffuse neutrino contribution from these sources which are compatible with
cosmic ray limits.  We shall also construct an upper bound for the
contribution of AGN jets, using the same method as in the previous section,
but for the appropriate generic source spectra, Eqs.~(\ref{eq.generic_cr})
and (\ref{eq.generic_nu}), varying the break energy $E_b$ over the range
allowed by the models.

\vspace{-1ex}
\subsection{Cosmic ray proton and neutrino production spectra from blazars}
\label{sect:pnu_blazar}
\vspace{-.5ex}

As our starting point, we assume a target photon spectrum with index
$\alpha=1$ which we have already seen leads to $\tau_{p \gamma}(E_p) \propto
E_p$.   A similar energy dependence applies to gamma rays interacting with the
same photons by photon-photon pair production ($\gamma\gamma\to e^+e^-$), and
so for $\alpha=1$ we have
\ifonecol\begin{equation}\else\begin{eqnarray}\fi
\label{eq.tau:pg:gg}
\tau_{p\gamma}(E_p) \ifonecol=\else&=&\fi 
	\frac{\langle{\kappa_p\sigma_{p\gamma}}\rangle}{
 	\langle\sigma_{\gamma\gamma}\rangle}\,
 	\tau_{\gamma\gamma}\!\left(\frac{2 m_e c^2 E_p}{
 	m_\pi [m_p+m_\pi/2]}\right) 
\ifonecol\approx\else\nonumber\\&\approx&\fi 
	5\mal 10^{-4}\, \tau_{\gamma\gamma}([4 \mal 10^{-6}] E_p)\;.
\ifonecol\end{equation}\else\end{eqnarray}\fi
Here we have used $\exval{\kappa_p\sigma_{p\gamma}} /
\exval{\sigma_{\gamma\gamma}} \approx 300\mybarn/\sigma_{\rm T}$ from
averaging over a photon spectrum with $\alpha=1$, where
$\exval{\kappa_p\sigma_{p\gamma}}$ includes Bethe-Heitler pair production,
$\sigma_{\gamma\gamma}$ is the total cross section for the process
$\gamma\gamma\to e^+e^-$ \cite{JauRoh}, and $\sigma_{\rm T}$ is the Thomson
cross section.  Using $\tau_{n\gamma}(E_n) \approx 2
\tau_{p\gamma}(E_n)$, and assuming that $\tau_{n\gamma}(E)\propto E$ holds
for a range $10^5 E_\gamma \lsim E_n \le E_{\rm max}$, we obtain the relation
\begin{equation}
\tau_{n\gamma}(E_n)/\tau_{\gamma\gamma}(E_\gamma) \approx 4
\times 10^{-9} E_n/E_\gamma.
\label{eq.tau_rel}
\end{equation}
We apply this relation to two reference AGN: the BL Lac object
Mrk\,501, and the quasar 3C\,279.  The combination of the
observed TeV spectrum and the EGRET flux limits for Mrk\,501
\cite{CBB+97apjl,KBB+99apj} gives rise to our assumption of a
break energy at about $0.3{-}1$~TeV in the context of
photohadronic models, while for 3C\,279 the EGRET spectrum
allows for a break at ${\sim} 3{-}10\GeV$ \cite{HWM+96apj}.  Note
that internal opacity due to the presence of low-energy
synchrotron photons with spectral index $\alpha$, which plays a
crucial role in photohadronic models for the gamma ray emission,
gives rise to a spectral steepening by $E^{-\alpha}$, and {\em
not} to an exponential cutoff, above the energy where
$\tau_{\gamma\gamma}=1$, see Eq.~(\ref{eq.pesc}). A TeV power law
being steeper than the gamma ray spectrum at EGRET energies could
thus imply optical thickness in spite of reaching some 25~TeV as
seen by the HEGRA telescopes \cite{AAB+97aa}.  Clearly, lower
values for the optical depth are obtained if the data are
interpreted with models which assume that the observed steepening
is due to the radiation process itself
(e.g. synchrotron-self-Compton emission)
~\cite{CBB+97apjl,KBB+99apj}.  Applying Eq.~(\ref{eq.tau_rel}) we
find that the break energy in the cosmic ray production spectra
should be $E_b \sim 10^{11}\GeV$ in Mrk\,501, and $E_b \sim
10^9\GeV$ in 3C\,279. Thus, UHE cosmic rays can escape from
Mrk\,501 under optically thin conditions, and our upper bound is
not affected by the actual value of the gamma ray optical depth
in this source.

We note that the break energies derived above depend on the assumption of an
undistorted power law target spectrum of photons.  This assumption might be
reasonable for BL Lacs, which show power law photon spectra extending from
the infrared (relevant for $p\gamma$ interactions) into the X-ray regime
(relevant for $\gamma\gamma$ absorption of ${\sim}10{-}100\GeV$ photons,
assuming a Doppler boosting of the emission by a factor $\delta\sim 10$).  In
high luminosity radio-quasars, however, this is not the case (see, e.g.,
\cite{GCFMC98mnras}), and relation (\ref{eq.tau_rel}) does not necessarily
hold.  Moreover, in these sources the dominant target spectrum for $p\gamma$
and $\gamma\gamma$ interactions could be given by the external accretion disk
photons, forming roughly an (unboosted) power law spectrum with $\alpha=1$ up
to ${\sim} 10\eV$ \cite{Pro97iau}, where it drops by about one order of
magnitude (e.g. \cite{MSR95aa}).  This still implies a relation like
Eq.~(\ref{eq.tau_rel}), but with $\tau_{n\gamma}$ increased relative to
$\tau_{\gamma\gamma}$ by a factor of $10$ --- consequently, the possible
break in the gamma-ray spectrum of 3C\,279 would correspond to $E_b\sim
10^8\GeV$.  A principal lower limit to the break energy in EGRET sources is
set at $E_b\sim 10^{7}\GeV$, since otherwise EGRET photons (${\gsim} 1\GeV$)
could not be emitted. 

Using cosmic ray proton production spectra with a break at ${\sim}
10^{8}\GeV$ (see Fig.~\ref{fig4}) will, of course, have a strong effect
on the neutrino bound implied by cosmic ray data.  Since the cosmic ray
proton spectrum of the source above the break drops faster than the
upper limit spectrum on  cosmic ray protons,
Eq.~(\ref{crmax}), the bound at a neutrino energy $E_\nu > E_b/25$ is
essentially set by the cosmic ray flux at the break energy $E_b$, rather
than by the more restrictive flux at $25 E_\nu$. If all sources had the
same $E_b$, the bound would be increased roughly by a factor $(25
E_\nu/E_b)^{0.75}$. Clearly, the assumption of optical depths allowing
break energies of $10^{8}\GeV$ or below in luminous quasars is not
directly supported, but only {\em consistent with} current observations.
The maximal neutrino intensity for AGN jets, 
which we shall derive  below on the basis of this assumption, is therefore at
the currently allowed maximum for the adopted source model, and likely to
be lowered when more detailed gamma-ray data become available.

As a general limitation of this approach it must be noted that most AGN jets
have not been detected by EGRET.  If their lack of high energy emission is
due to a large gamma-ray opacity, this would also imply very low break
energies for their ejected cosmic ray spectrum. One such possible class of
sources are the gigahertz-peaked sources (GPS) and compact steep-spectrum
(CSS) quasars~\cite{ODea98}, which make up about 40\% of the bright radio
source population.  Objects such as these could thus produce a diffuse
neutrino intensity at the level of the EGRB or maybe even above, without
violating any constraint.  Also low-opacity sources for which the maximum
proton energy remains much less than its theoretically allowed value and, in
particular, below the value of the break energy, could lead to a higher
neutrino intensity at low $E_\nu$ (limited only by the EGRB).  We did not
consider this possibility by keeping $E_{p,{\rm max}}$ fixed.
%
%
\def\captionfigfour{Form of the spectra of neutrinos and cosmic
rays escaping from optically thick photoproduction sources,
Eqs.~(\ref{eq.generic_cr}) and (\ref{eq.generic_nu}).  The example shown here
has a break energy of $10^8$ GeV, and an exponential cut-off above
$10^{11}\GeV$.}
\ifsubmit\else
\begin{figure}[t]
\centerline{\kern0.4cm\epsfysize7.1cm\epsfbox{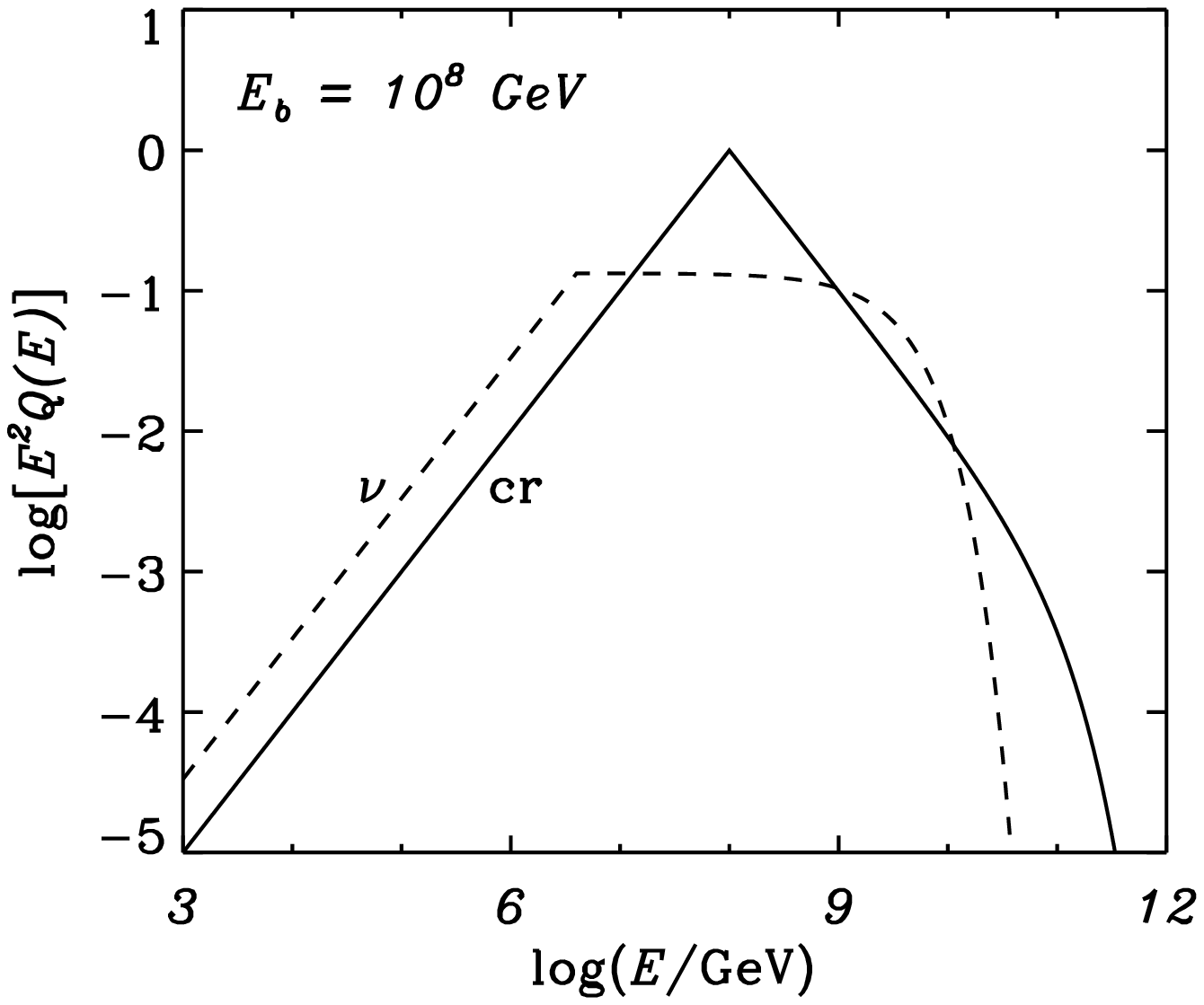}}
\caption[]{\label{fig4}\capsize\captionfigfour}
\end{figure}
\fi 
%

\vspace{-1ex}
\subsection{Blazar luminosity functions and generic models for the neutrino
contribution from BL Lacs and radio-quasars}
\vspace{-.5ex}

In order to obtain a parametrization of blazar neutrino spectra, we need to
express $\tau_{n\gamma}$ as a function of the blazar luminosity $L$, given at
some frequency.  Here we take into consideration that blazars are assumed to
be beamed emitters, with a Doppler factor $\delta\sim 10$.  The optical depth
intrinsic to the emission region, is proportional to $L_{\rm t}/R$, where
$L_{\rm t}$ is the intrinsic target photon luminosity and $R$ is the size of
the emitter.  Since blazars are strongly variable objects \cite{UMU97araa},
we can use the variability time scale, $T_{\rm var}$, to estimate the
intrinsic size by $R\sim T_{\rm var} c \delta$.  Using also the relation between
intrinsic and observed luminosity, $L = L_{\rm t} \delta^4$, we obtain
$\tau_{n\gamma}\propto L\,T_{\rm var}^{-1} \delta^{-5}$.  Although there is no
detailed study of a possible systematic dependence of $T_{\rm var}$ on $L$,
the observations are compatible with no such correlation existing --- for
example, variability time scales of order 1 day are common in both,
moderately bright BL Lacs like Mrk 501 or Mrk 421, and powerful quasars like
3C\,279, PKS\,0528+134, or PKS\,1622-297, whose optical luminosities differ
by at least three orders of magnitude \cite{UMU97araa}.  For the Doppler
factors, there is no evidence for a systematic dependence on $L$ either
\cite{GD97apj}, although unification models for blazars and radio galaxies
(see next section) suggest that BL Lacs are on average slightly less beamed
($\exval{\delta}\sim 7$) than radio-quasars ($\exval{\delta}\sim 11$)
\cite{UPad95pasp}.  Therefore, we may assume that for both BL Lacs and
quasars, $\tau_{n\gamma}(E,L) \propto L$ holds on average, and that
$\tau_{n\gamma}(E,L,{\rm BL \, Lac}) \sim 10 \tau_{n\gamma}(E,L,{\rm
quasar})$.  It is interesting to note that this relation would imply a
relation of the break energies of Mrk\,501 ($L_{\rm opt}\sim
10^{44}\erg\scnd^{-1}$) and 3C\,279 ($L_{\rm opt}\sim 10^{47}\erg/\scnd$) as
$E_b({\rm 3C\,279}) \sim 10^{-2} E_b({\rm Mrk\,501})$, consistent with our
estimate obtained for intrinsic absorption from the gamma ray spectral break.
Of course, this does not rule out the possibility that $E_b$ might be
systematically lower in 3C\,279 due to external photons, as argued above.

To determine the contribution of all blazars in the Universe, we have to
relate the proton luminosity $L_p$ to the blazar luminosity $L$ in some
frequency range, and then integrate over the luminosity function, $d\cN/dL$,
determined for the same frequency range.  Here we have to distinguish between
the luminosities in the energy range where the target photons are, $L_o$, and
the luminosity of the gamma-rays, $L_\gamma$, which are here assumed to be
produced by hadronic interactions.  Obviously, the latter implies $L_p\propto
L_\gamma$, but on the other hand $\tau_{n\gamma}\propto L_o$.  We also have
to distinguish between the two classes of blazars: while for BL Lacs
$L_\gamma \propto L_o$, observations rather suggest that for quasars
$L_\gamma \propto L_o^2$ \cite{GCFMC98mnras}.  This leads to the following
simple models:

For BL Lacs, we use the X-ray luminosity functions given by Wolter et al.\
\cite{WCDM94apj} for X-ray selected BL Lacs,
\ifonecol\begin{equation}\else\begin{eqnarray}\fi
\ifonecol{\else\lefteqn{\fi
d\cN_{\rm BL}/dL_{\rm X} \propto L_{\rm X}^{-1.6}}
\ifonecol\relax\else\\&&\fi 
	\quad (3\mal 10^{43} \erg\scnd^{-1} <L_{\rm X}<
3\mal 10^{46} \erg\scnd^{-1}), 
\ifonecol\relax\else\nonumber\fi
\ifonecol\end{equation}\else\end{eqnarray}\fi
and use the relations $L_p \propto L_{\rm X}$, and $E_b\propto L_{\rm
X}^{-1}$ with $E_b = 10^{11}\GeV$ for $L_{\rm X} = 3\mal
10^{44}\erg\scnd^{-1}$ (Mrk\,501).  Here we have assumed that $L_o\propto
L_{\rm X}$.

For quasars, we use the EGRET luminosity function given by
Chiang \& Mukherjee \cite{CM98apj},
\ifonecol\begin{equation}\else\begin{eqnarray}\fi
\ifonecol{\else\lefteqn{\fi
d\cN_{\rm q}/dL_\gamma \propto L_\gamma^{-2.2}}
\ifonecol\relax\else\\&&\fi 
	\quad (10^{46}\erg\scnd <L_\gamma< 10^{48} \erg\scnd^{-1}), 
\ifonecol\relax\else\nonumber\fi
\ifonecol\end{equation}\else\end{eqnarray}\fi
and the relations $L_p\propto L_\gamma$, and $E_b\propto L_\gamma^{-1/2}$
where we consider two possible normalizations for $L_\gamma =
10^{48}\erg\scnd^{-1}$ (3C\,279), which are $E_b = 10^9\GeV$ in case that
jet-intrinsic photons dominate the target field, and $E_b = 10^8\GeV$ for the
assumption that external photons are the dominant target.  For illustration,
we show in Fig.~\ref{fig4} cosmic ray and neutrino spectra on emission for
$E_b=10^8\GeV$ and $E_{\rm max}=10^{11}\GeV$.

Then we obtain the form of the production spectra of cosmic ray protons and
neutrinos averaged over the local universe due to these two classes of AGN,
\begin{equation}
\langle Q_{cr,\nu}(E) \rangle = {\int Q_{cr,\nu}(E,L) (d\cN/dL) dL 
\over \int L \, (d\cN/dL) dL},
\end{equation}
where the input spectra $Q_{\rm cr}$, $Q_\nu$ are given by
Eqs.~(\ref{eq.generic_cr}) and (\ref{eq.generic_nu}), with $E_{\rm max} =
10^{11}\GeV$, and $E_b$, $L_p$ given as functions of $L$ as discussed above.

To integrate properly over redshift, we note that while quasars
show strong evolution similar to galaxies (see below), BL Lacs
show little or no evolution \cite{BBD+98aa}, and we shall take this into
account when propagating these spectra through the Universe from
large redshifts.   For quasars, their luminosity per co-moving
volume has a pronounced peak at redshifts of $z\sim 2$, and
declines or levels off at higher redshifts \cite{MC99app}.  We
shall assume that this effect is due to evolution of the number
of quasars with $z$, rather than evolution of the luminosity of
individual sources, which keeps the production spectra $\langle
Q_{{\rm cr},\nu}(E) \rangle$ independent of $z$.  A particular
parametrization of the redshift-dependence of the (co-moving
frame) UV luminosity density of AGNs as inferred by Boyle and
Terlevich \cite{BT98mnras}, assuming an Einstein-de Sitter cosmology and
$h_{50}=1$, is given by
\begin{equation}
{dP_{\rm gal} \over dV_c} = P_0 \left\{ 
\begin{array}{ll} 
        [(1+z)/2.9]^{3.4}  & (z < 1.9) \\
        1.0                & (1.9 \le z < 3)\\
        \exp[-(z-3)/1.099] & (z \ge 3)
        \end{array}\right.\quad,
\label{eq.boyle}
\end{equation}
where $P_0 = (3.0 \pm 0.3)\mal 10^{44}\erg$~s$^{-1}$~Mpc$^{-3}$.  Clearly, the
normalization plays no role here since $L_p$ is adapted to match the cosmic
ray flux at earth.  So for BL Lacs (no evolution), we simply use $dP_{\rm gal}
/ dV_c=1$.

The result is shown in Figure~\ref{fig5}.  As expected, the diffuse neutrino
fluxes from AGN jet models exceed the bound for optically thin sources, but
fall into the allowed region for sources optically thick for neutron
emission.  The BL Lac contribution falls below the bound, because it was
derived for a non-evolving source distribution --- we note that it still
exceeds the corresponding Waxman-Bahcall bound for the case of no evolution.
In addition to the models discussed above, we have also constructed and
estimated an upper bound on the diffuse neutrino contribution on AGN jets.
Here we used the same method as for the construction of Fig.~\ref{fig3}, but
implying input spectra of the form of Eqs.~(\ref{eq.generic_cr}) and
(\ref{eq.generic_nu}), with variable $E_b$ and fixed $E_{\rm max} =
10^{11}\GeV$.  The break energy $E_b$ was then varied in the range
$10^7\GeV<E_b<E_{\rm max}=10^{11}\GeV$, and the normalization chosen such
that the superposed spectra approximately represented the upper limit on the
extragalactic cosmic ray spectrum (Fig.~\ref{fig5}, left panel).  We note
that this upper bound corresponds within a factor of $2$ with the prediction
of a previously published model by Protheroe \cite{Pro97iau}.  Other
published models, for example by Halzen \& Zas \cite{HZ97apj}, or Mannheim
\cite[Model B]{Man95app}, exceed our bound by about one order of magnitude at
energies $E_\nu \gsim 10^8\GeV$, but their predictions for the important
energy range below ${\sim} 10^{7}\GeV$ are compatible with this bound.  We
return to the discussion of these models later, after we discuss the effect
of magnetic fields in the AGN environment in the next section.
%
%
\def\captionfigfive{(a) Comparison of neutrino spectra for optically thick
pion photoproduction sources with neutrino upper bounds obtained in the
present work and by Waxman and Bahcall \protect\cite{WB99prd}.  Neutrino
intensities obtained in the present work plotted are: maximal superposition
of spectra having $E_b$ in the range $10^7-10^{11}\GeV$ (thick solid curve);
maximum source spectra averaged over the quasar luminosity function assuming
$\tau_{n\gamma} \propto L^{1/2}$ with break energies corresponding to
$L=10^{48}$~erg s$^{-1}$ at $E_b\approx10^8\GeV$ (thick dashed curve) and at
$E_b\approx10^9\GeV$ (thick dotted curve); maximum source spectra averaged
over the BL Lac luminosity function assuming $\tau_{n\gamma} \propto L$ with
the break energy corresponding to $L=3 \times 10^{44}$~erg s$^{-1}$ at
$E_b\approx10^{11}\GeV$ (thin dot-dashed curve).  Also shown is the
prediction by Protheroe \protect\cite{Pro97iau} for an external photon
optically thick proton blazar model normalized down in accordance with the
recent estimates of the blazar contribution to the diffuse gamma ray
background \protect\cite{MC99app} (thick chain curve), and the bounds
obtained by Waxman \& Bahcall \protect\cite{WB99prd} with and without source
evolution (note that the maximum BL Lac intensity was calculated assuming no
evolution).  Other symbols shown correspond to Figure~\ref{fig3}.  (b) Cosmic
ray intensities for the neutrino intensities obtained in the present work
shown in part a (key to these curves is as in part a).  The upper bound to
any extragalactic cosmic ray proton spectrum is shown by the thin line, which
corresponds to the dotted line in Figure~\protect\ref{fig1}.}
\ifsubmit\else
\ifrevtex\begin{figure*}[t]\else\begin{figure}[tp]\fi 
\centerline{\kern0.4cm\epsfysize7.1cm\epsfbox{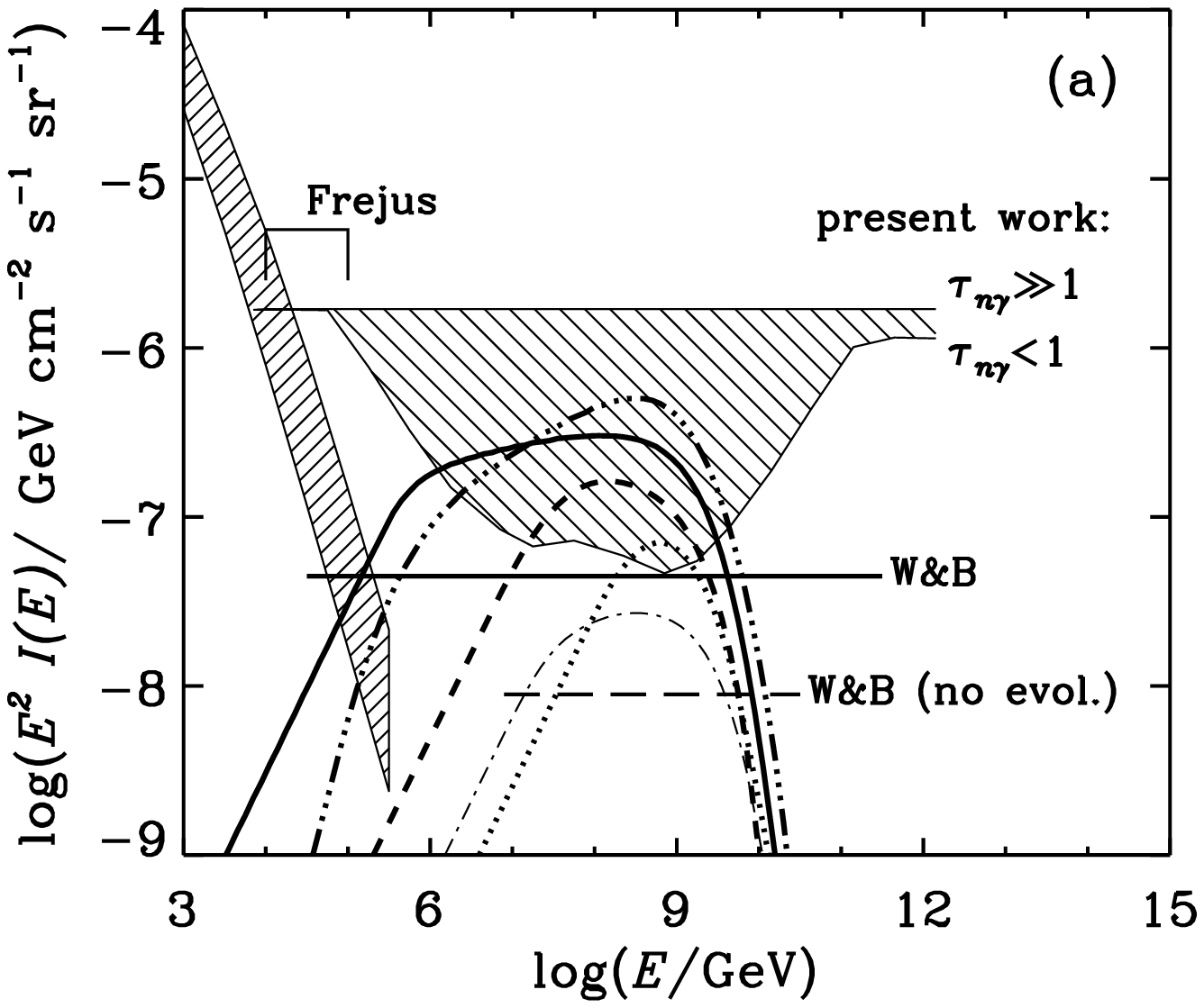}\hss
        \epsfysize7.1cm\epsfbox{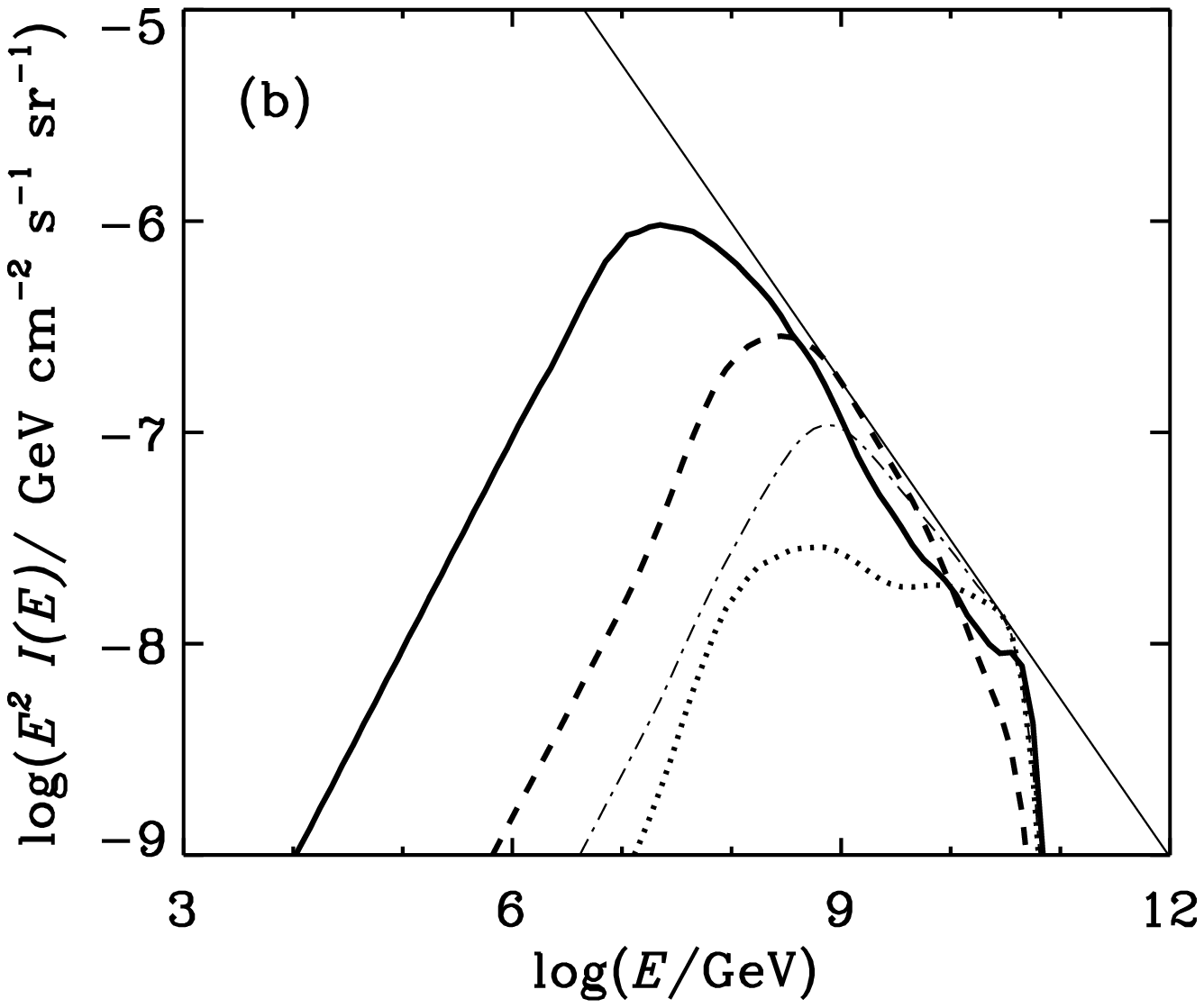}} 
\caption[]{\label{fig5}\capsize\captionfigfive} 
\ifrevtex\end{figure*}\else\end{figure}\fi \fi 

\section{Magnetic fields and their impact on cosmic ray propagation}
\label{sect:magnetic}

As discussed in Sect.\,II, protons accelerated inside AGN jets
are likely to be trapped in the jet to be released later near the
end of the jet, and will consequently suffer severe adiabatic
losses as a result of jet expansion.  The bound that we
calculated for optically thin photoproduction sources shown in
Fig.~3 may of course be exceeded for optically thick sources.
However, it may even be exceeded for optically thin
photoproduction sources if the cosmic rays resulting from
photoproduced neutrons are trapped, or suffer severe adiabatic
deceleration in the large-scale magnetized environment such as the host
galaxy and its halo, a galaxy cluster, or a supercluster.  We
emphasize that although there is now useful information
available about magnetic fields in galaxies and clusters, the
magnetic field structure and topology is not sufficiently well
known for us to predict reliably magnetic trapping and adiabatic
losses in host galaxies and clusters.  However, we shall discuss in
this section the fate of the cosmic rays resulting from
photoproduced neutrons, and show that in some plausible
scenarios, our bound for optically thin photoproduction sources
can be exceeded for neutrinos from optically thin photoproduction
sources.     

X-ray observations and measurements
of extragalactic Faraday rotation suggest that structures
surrounding compact AGN jets
carry magnetic fields of the order of $0.1$-$10\muG$
\cite{Kron94rpp}.  These can influence the propagation of cosmic
ray protons in essentially two ways: (a) particles may be
physically confined in the structure for a time $t\ge t_{\rm H} =
1/H_\circ$, or (b) the diffusive escape of the particles can lead
to adiabatic energy losses.  Obviously, magnetic fields on scales
larger than the mean free path of a neutron for $\beta$-decay,
$l_n \approx 10\kpc (E_n/10^{9}{\rm \, GeV})$, also affect cosmic
rays which are ejected as neutrons from the source.  Here we
shall discuss what influence these effects can have on the
strength of the measured cosmic ray flux relative to the
corresponding neutrino flux.  We shall estimate the critical
energy $E^*$ below which our bound could plausibly be exceeded
for optically thin photoproduction sources for a number of
scenarios, in particular for clusters of galaxies and radio
galaxies hosting AGN.

\subsection{Particle confinement in clusters and superclusters}

Clusters of galaxies have been recently discussed in the literature as a
possible ``storage room'' for cosmic rays \cite{VAB96ssr,BBP97apj,EBKW97apj},
because of their relatively strong magnetic fields ($B_0 \gsim 1\muG$)
extending over large scales, i.e.\ cluster radii, $R_{\rm cl} \gsim 1\Mpc$.
The time scale for diffusive escape of cosmic rays in a turbulent magnetic
field of homogeneous strength $B_0$ within a central radius $R_0$ is given by
\begin{equation}
\label{tesc:D}
t_{\rm esc} \sim \frac{R_0^2}{D} = 
	\frac{3 e B_0 R_0^2}{c E_{\rm cr} \lambda(E_{\rm cr})}\;, 
\end{equation}
where we have used for the diffusion coefficient $D = \frac13 \lambda r_{\rm
L} c$, and $\lambda(E_{\rm cr})$ is the scattering length in units of the
Larmor radius, $r_{\rm L} = E_{\rm cr}/e B_0$ of a cosmic ray proton of
energy $E_{\rm cr}$.  The function $\lambda$ depends on the turbulence
spectrum of the magnetic field, $[\delta B(k)]^2 \propto k^{-y}$, where $k$ is
the wavenumber.  Cosmic ray scattering is dominated by field fluctuations
on the scale of the Larmor radius, i.e., $k\sim r_{\rm L}^{-1}$.  This
implies that
\begin{equation}
\label{lambda}
\lambda(E_{\rm cr}) = \left[\frac{e B_0 a_0}{E_{\rm cr}}
                                \right]^y \for E_{\rm cr}<e B a_0\;, 
\end{equation}
where $a_0 = k_{\rm min}^{-1}$ is the largest scale of the turbulence, i.e.,
the ``cell-size'' (or ``reversal-scale'') of the magnetic field.  For
$k>a_0^{-1}$, the magnetic turbulence in clusters of galaxies seems to be
well described by the Kolmogorov law for fully developed hydrodynamical
turbulence, i.e., $y=\frac23$.  This can be easily seen from relating the
typical turbulent magnetic field and cell size found from Faraday rotation
measurements, $a_0 \sim 20\kpc$ and $B_0\sim 1\muG$ \cite{DCP87apj,KTK91apj},
to the diffusion coefficient found for electrons of energy ${\sim}1\GeV$ from
the synchrotron radio emission spectrum \cite{SST87aa}, $D\sim 2\mal
10^{29}\cm^2\scnd^{-1}$.  For $E_{\rm cr} > e B a_0$, i.e.\ $r_{\rm L}>a_0$,
the particle motion is a random walk with scattering angles ${\sim}
a_0/r_{\rm L}$ in each step. This can also be approximately described by
Eqs.~(\ref{tesc:D}) and (\ref{lambda}) by setting $y=-1$.

Confinement of cosmic rays over the cluster radius, $R_0 = R_{\rm cl} \sim
3\Mpc$, is then obtained for $t_{\rm esc} > t_{\rm H}$, corresponding to a
critical energy
\begin{equation}
\label{Econf:diff}
E_{\rm diff}^* \approx 5\mal 10^{8}\,{\rm GeV} \times
\left[\frac{B_0}{1\muG}\right] \left[\frac{R_{\rm cl}}{3\Mpc}\right]^6  
\left[\frac{a_0}{20\kpc}\right]^{-2},
\end{equation}
provided $a_0 > 1\kpc (R_{\rm cl}/3\Mpc)^2$.  This assumes
that the magnetic field strength is homogeneous over the entire
cluster.  However, if it dropped to $\sim 0.1 \mu$G, and $a_0
\sim 200$~kpc at the edge of the cluster then $E_{\rm diff}^*$
would be a factor ${\sim}10^3$ lower.  On the other hand, some
observations suggest also larger magnetic fields on smaller
reversal scales \cite{FDGT95aa,FDG+99aa}, which would imply
higher confinement energies.

The  scenario above assumes that the background plasma filling the
cluster is at rest.  However, simulations of structure formation suggest 
that clusters of galaxies are accreting extragalactic hot gas
\cite{KCOR94apj}, forming inflows of typical speeds $v_{\rm acc}\gsim
300\km\scnd^{-1}$ downstream of an accretion shock near the outer radius of
the cluster \cite{KRJ96apj}.  Since particles diffuse relative to the plasma
flow, we require that in order to escape from the cluster, the average
``speed'' of particle diffusion, ${\sim}D/R_{\rm cl}$, exceeds
the inflow velocity.  If we assume the lower value for the
magnetic field at $R_{\rm cl}\sim 3\Mpc$, i.e.  $B_0\sim 0.1\muG$
and $a_0\sim 200\kpc$, we obtain a critical energy
\ifonecol\begin{equation}\else\begin{eqnarray}\fi
\ifonecol{\else\lefteqn{\fi E^*_{\rm acc} \approx 2\mal 10^{6}\GeV \times}
\ifonecol\relax\else\\&&\quad\fi \left[\frac{B(R_{\rm
cl})}{0.1\muG}\right]
\left[\frac{R_{\rm cl}}{3\Mpc}\frac{v_{\rm
acc}}{300\km\scnd^{-1}}\right]^3
\left[\frac{a(R_{\rm cl})}{200\kpc}\right]^{-2}.
\ifonecol\relax\else\nonumber\fi
\ifonecol\end{equation}\else\end{eqnarray}\fi
Of course, with higher magnetic fields and a shorter
field-reversal length $a_0$, $E^*_{\rm acc}$ would be higher.
Similar processes would occur in the cooling flows observed in
some rich clusters harboring powerful radio galaxies, but the
confinement energies for typical cooling flow parameters
\cite{WS87apj3,TP93apj} are even lower. Also the effect of drift
in the nonhomogeneous cluster field \cite{WB99prd} is small compared to
diffusion, and can not lead to larger confinement energies.

Ensslin et al.\ \cite{EBKW97apj} point out that simple diffusion, as
assumed above, may not be the best description of cosmic ray 
propagation in clusters of galaxies, and suggest that a one-dimensional
random walk along static, but randomly tangled, magnetic field lines may be
more realistic for particles with $r_{\rm L}\ll a_0$.  In this case the field
line topology may be considered as arising from a three-dimensional random
walk with steps of size $a_0$, yielding an effective diffusion length of $R_0
= R_{\rm cl}^2/a_0$. For the escape time we then obtain
\begin{equation}
t_{\rm esc} = \frac{R_{\rm cl}^4}{2 c a_0^3} 
        \left[\frac{e B_0 a_0}{E_{\rm cr}}\right]^{1/3} \gsim t_{\rm H} 
        \for E_{\rm cr} \ll e B_0 a_0\;,
\end{equation} 
which means for typical cluster parameters that essentially all cosmic
rays with $E_{\rm cr} \lsim 3 \mal 10^9\GeV$ are confined to the cluster. Of
course, this result neglects cross-field diffusion so that the maximum
confinement energy is probably lower, but it could still be higher than the
result obtained in Eq.~(\ref{Econf:diff}) for the case of simple diffusion,
in particular in the likely case that the field strength decreases with
the distance from the cluster core.

Confinement of cosmic rays in clusters would lead to a decrease of the cosmic
ray flux measured at earth relative to the corresponding neutrino flux,
causing an increase of the neutrino bound.  It is important to note, however,
that cosmic ray confinement may exist even on larger scales, i.e.,
superclusters.  It has been shown that magnetic fields $\sim 0.1\muG$ in
superclusters are consistent with observations, and expected in simulations
of structure formation which also predict accretion of gas with speeds
$v_{\rm acc}\sim 1000\km\scnd^{-1}$ \cite{BKRR97moriond,RyuB98aa}.  The
cosmic ray confinement energies for these larger scales ($R \sim 10\Mpc$,
$a\sim 1\Mpc$) could therefore be ${\gsim}10^8\GeV$, where again topological
aspects connected to the detailed structure of the field may allow even
higher values.  Since our Galaxy itself is located in a supercluster, this
latter scenario would tend to {\em increase} the cosmic ray flux relative to
the corresponding neutrino flux, which fills the Universe homogeneously ---
thus it could actually decrease the bound below $E_\nu\sim 10^7\GeV$. The net
effect of confinement in clusters and our local supercluster is obviously
strongly model dependent, and therefore difficult to estimate.

An effect of the unknown structure of magnetic fields on galaxy cluster and
supercluster scales may therefore most likely affect the observed
extragalactic cosmic ray flux at energies below ${\sim} 2\mal 10^6\GeV$,
corresponding to $E_\nu\sim 10^5\GeV$. On the other hand, we can exclude an
effect only for energies $E_{\rm cr}\gsim 3\mal 10^9\GeV$ ($E_\nu\gsim
10^8\GeV$), but we note that cosmic ray confinement at energies higher than
$\sim 10^8\GeV$ requires extreme assumptions on the strength and topology of
the magnetic fields. We therefore agree with the conclusion of Waxman \&
Bahcall \cite{WB99prd} that at $E_\nu\sim 5\mal 10^8\GeV$ an effect of large
scale magnetic fields on the relation of cosmic ray and neutrino fluxes
cannot be expected.

\vspace{-1ex}
\subsection{Adiabatic losses in expanding radio lobes and halos}
\vspace{-.5ex}

Although many galaxy clusters have powerful radio galaxies
at their centers, it is also a fact that most powerful radio galaxies are not
found in such environments \cite{LS79mnras,Sto79apj}.  For a radio galaxy
located in the normal extragalactic medium, evidence has been found that a
pressure equilibrium with the external medium cannot be obtained within the
lifetime of the source (${\lsim} 10^8\yr$) \cite{MLFTE85mnras}.  Therefore,
it can be concluded that the lobes must expand.  For a sample of powerful
double-lobe (or FR-II) radio galaxies, lobe propagation velocities of order
$10^4\km\scnd^{-1}$ have been inferred \cite{Daly95apj}.  Since the aspect
ratio of the sources is found to be independent of their size, consistent
with a propagation of the lobes along a constant opening angle of ${\sim}
10\deg$, expansion velocities of the lobes and the connecting ``bridges'' of
$v_{\rm ex} \sim 1000\km\scnd^{-1}$ can be inferred.

Neutrons produced by pion photoproduction interactions at an acceleration
site near the base of the jet will be beamed preferentially along the jet
direction decaying farther out along the jet or in the radio lobe, where they
decay.  The resulting cosmic ray protons will then be advected with the
outflowing plasma on a time scale of the galaxy life-time ${\sim} 10^8\yr \ll
t_{\rm H}$. Additionally, the protons perform a random walk in the magnetic
field, which may even decrease their confinement time.  Expanding radio lobes
can therefore not confine cosmic rays indefinitely.  However, in plasma
outflows all particles which are isotropized in the flow due to scattering
with plasma turbulence will experience adiabatic losses before their release.
Kinetic theory implies that the particle energy at ejection from the flow is
related to the injected particle energy by
\begin{equation}
\label{adloss}
\frac{E_{\rm ej}}{E_{\rm inj}} = \frac{R_{\rm inj}}{R_{\rm max}}\;.
\end{equation}
Here, $R_{\rm max}$ is the radius of the outer termination shock
of the flow, and $R_{\rm inj}$ is in general given by some
minimum radius $R_{\rm min}$ where the outflow starts.  For the
case of neutron ejection from a central source, as considered
here, $R_{\rm inj} = l_n$ if $l_n > R_{\rm min}$, and $R_{\rm
inj} = R_{\rm min}$ otherwise.  Note that the lobes we are
considering here are much more extended than the jets in which
the particles are accelerated.

The observed synchrotron spectra from radio lobes imply typical
magnetic fields in the range $10{-}50\muG$, turbulent with a
maximum scale (or cell-size) of ${\sim} 0.5\kpc$
\cite{RHP90aa,Daly95apj}.  The observed asymmetric depolarization
in double radio galaxies leads to the suggestion that the
magnetized plasma around the radio galaxy extends into a halo of
radius ${\sim} 300\kpc$, with $B\sim 0.3\muG$ and a cell size of
${\sim} 5\kpc$ at a radius $R\sim 100\kpc$ \cite{GC91mnras}.  The
properties of the magnetic field in the central lobe and the halo
can be connected by assuming that magnetic field and cell size
scale as $B = B_0 (R/R_0)^{-2}$ and $a = a_0 (R/R_0)$,
respectively, with $B_0 \sim 30\muG$, $a_0 \sim 0.5\kpc$ and $R_0
= 10\kpc$.  

Within radius $R_0$, we assume the properties of the turbulent magnetic field
are constant.  This corresponds to the assumption of an isotropic magnetic
field expanding in a plasma outflow with $R_{\rm min} = R_0$ and $R_{\rm max}
\sim 300\kpc$, which will be used as a working hypothesis in the following.
We can consider cosmic ray protons as nearly isotropized in the plasma if
$r_{\rm L}(R) < a(R)$, corresponding to energies $E_{\rm cr}(R)<E^*_{\rm
ad}(R)$ with
\begin{equation}
E^*_{\rm ad}(R) = 10^{10}\GeV \times \left[\frac{B_0}{30\muG}\right]
\left[\frac{a_0}{0.5\kpc}\right] \left[\frac{R}{10\kpc}\right]^{-1}
\end{equation}
for $10\kpc < R \lsim 300\kpc$. To justify the assumption of near-isotropy
below $E^*_{\rm ad}(R)$, we have to show that advection in the flow indeed
dominates over diffusion, i.e., that the centers of the diffusive cloud of
cosmic ray protons move approximately steady with the flow. Indeed, for the
parameters adopted above we find for the ``speed'' of diffusive escape $D/R
\sim \frac13 a(R) v_{\rm cell}/R < v_{\rm ex}$, where $v_{\rm cell}\le
\frac13c$ is the average velocity of the particles to cross the cell,
considering that also this motion is diffusive if the magnetic turbulence
proceeds to smaller scales as expressed in Eq.~(\ref{lambda}).  

Consequently, we can consider the cosmic rays with $E_{\rm cr} \lsim E^*_{\rm
ad}(R)$ as are advected with the flow at a distance $R$ from the center of
the galaxy, so that they suffer adiabatic losses following
Eq.~(\ref{adloss}) leading to $E_{\rm cr}(R) \propto R^{-1}$.  This implies
that $E_{\rm cr}/E^*_{\rm ad}$ is independent of $R$, i.e. that cosmic rays
confined in the outflow at some radius $R$ remain confined for larger radii.
If we consider cosmic rays which are ejected as neutrons, the radius where
they couple to the magnetic field is given by the $\beta$-decay mean free
path, $l_n \approx 10\kpc (E_n/10^{9}\GeV)$.  The resulting protons are
confined to the outflow and are subject to adiabatic losses provided $E_n <
E^*_{\rm ad}(l_n)$, or
\begin{equation}
E_n < 3\mal 10^{9}\GeV \left[\frac{B_0}{30\muG}\right]^{1/2}
\left[\frac{a_0}{0.5\kpc}\right]^{1/2}
\end{equation}
The energy on ejection from the outflow is then $E_{\rm ej} \approx E_n/30$
for $E_n<10^9\GeV$, and $E_{\rm ej} = E_n^2/(3\mal 10^{10}\GeV)$ for
$10^{9}\GeV < E_n < 3\mal10^9\GeV$.  For $E_n > 3\mal 10^9\GeV$, cosmic rays
(neutrons and protons) traverse the lobe/halo along almost straight paths and
adiabatic losses do not apply. (Note that the corresponding energies where
these modifications may affect the neutrino bound are $E_\nu\sim E_n/20$,
thus $E_\nu<10^8\GeV$.)

Energy losses of cosmic rays in the lobes and halos of radio galaxies are of
particular relevance for models of neutrino production in AGN jets, which we
discussed in Sect.\,\ref{sect:agn}.  These models apply to radio loud AGN,
which are likely to be the beamed counterparts of radio galaxies
\cite{UPad95pasp}.  The two classes of AGN discussed in the last section
correspond to the two Fanaroff-Riley (FR, \cite{FR74mnras}) classes of
radio galaxies: radio quasars might be associated to the powerful
double-lobed FR-II radio galaxies, while BL Lac objects might correspond to
the less luminous FR-I radio galaxies which generally have diffuse lobes
centered around the AGN.  The parameters used above for the lobes/halos of
radio galaxies were mainly obtained from observations of FR-II radio galaxies
or radio quasars.  Therefore, the cosmic ray ejection from radio quasar/FR-II
sources can be expected to be diminished by more than an order of magnitude
below ${\sim} 10^9\GeV$.  For the less luminous FR-I galaxies, it could be
that the magnetic fields, turbulence scales and halo sizes used above are
overestimated, so that $E^*_{\rm ad}$ could be lower by about one or two
orders of magnitude for these sources.

\section{Discussion and conclusions}

We have looked at the problem of to what extent a possible flux of
extragalactic neutrinos in the broad energy range
$10^5\GeV\,{\lsim}\,E_\nu\,{\lsim}\,10^{11}\GeV$ is bounded by the observed
flux of cosmic rays and gamma rays.  As the minimum contribution to the
cosmic rays we consider the neutrons produced together with other neutrals
such as gamma rays and neutrinos in photohadronic interactions.  The most
restrictive bound arises for sources which are transparent to the emission of
neutrons and where the protons from the decaying neutrons are unaffected by
large-scale magnetic fields.  The bound for this case is approximately in
agreement with the bound previously computed by
Waxman\,\&\,Bahcall\cite{WB99prd} in the neutrino energy range
$10^7\GeV\,{\lsim}\,E_\nu\,{\lsim}\,10^9\GeV$ but is higher at lower and at
higher energies. The difference is a consequence of the different approaches:
Waxman\,\&\,Bahcall assume a {\em particular model spectrum}, which could
explain the observed UHECR spectrum, but which has a pronounced GZK cutoff
that is not clearly confirmed by the present low statistics data, and which
does not reproduce the steep slope of the observed CR spectrum below
${\sim}10^{10}\GeV$. It thus leaves room for additional contributions from
extragalactic sources (with different spectra) outside the narrow energy
range where it matches with the observed cosmic ray flux. Our approach
considers this possibility by using the best currently available {\em
experimental} upper limit on the extragalactic proton contribution.  We have
also pointed out in Sect.\,\ref{CGN_generic}, referring to recent
Monte-Carlo simulations\cite{SOPHIA99texas}, that the fundamental properties
of photohad\-ronic interactions can affect the bound by up to a factor of\,$5$.

At neutrino energies below $10^7\GeV$, the cosmic ray bound computed in this
work rises and equals the bound inferred from the EGRB flux at about
$10^5\GeV$, below which the EGRB constraint is tighter.  In the same energy
region, we have also shown that effects from extragalactic magnetic fields
come into play; they could either increase or reduce the bound, depending on
the details of the field strength and structure, and the distribution of
cosmic ray/neutrino sources.  We therefore do not regard the bound based on
the cosmic ray flux as a firm upper limit on neutrino fluxes below about
$10^5\GeV$, the main energy region for underwater/ice Cherenkov experiments
such as AMANDA \cite{Hal98nar}.  Models which predict neutrino emission
mainly in this energy range are therefore not rigorously bounded by cosmic
ray data, as for example the model of Berezinsky et al.\ \cite{BBP97apj}
predicting neutrino fluxes from cosmic rays stored in clusters of galaxies.
Such sources could in principle produce neutrino fluxes almost up to the EGRB
limit within the AMANDA range.  However, the neutrino contribution from
clusters of galaxies is expected to be much lower, as it is limited by their
expected contribution to the EGRB of ${\sim}1\%$ \cite{CB98app}.

At neutrino energies above $10^{9}\GeV$, the upper limit rises up to the
point where the energy flux of secondary gamma rays increases above the level
of the observed EGRB.  The reason for the increasing bound is that while
cosmic rays from evolving extragalactic sources above the nominal GZK cutoff
reach us exponentially damped due to interactions with the microwave
background, neutrinos reach us essentially unattenuated.  An observed excess
of cosmic ray events above the GZK cutoff would therefore correspond to an
even more pronounced excess of neutrinos in the framework of such models.
Cosmic rays from a fiducial class of extragalactic photohadronic sources with
a sufficiently flat spectrum could thus saturate the rising bound at the
highest energies and would still remain below the local cosmic ray flux at
all energies.  Clearly, this is not the most likely scenario to explain the
events beyond the GZK cutoff. For example, if these events are due to a
single strong nearby source, then we would not expect the extragalactic
neutrino flux above $10^{9}\GeV$ to be at a level higher than given by the
bound computed by Waxman\,\&\,Bahcall, even if this source would have the
spectral properties which we used to construct our bound in
Sect.\,\ref{sect:thinbound}. On the other hand, we note that the flux from
non-photohadronic sources, as for example from the decay of topological
defects, can exceed this bound because of their different branching ratios
for the production of baryons relative to mesons \cite{PS96prl}. Measuring
the neutrino flux in this energy region, as is planned using large air-shower
experiments such as the Pierre Auger Observatory \cite{CCPZ98app}, would
therefore be highly relevant for understanding the nature and cosmic
distribution of UHECR sources.

The bound we derive for extragalactic neutrinos is indeed an upper limit,
since we do not consider the possibility that ultrarelativistic protons are
ejected from their sources without preceding isospin flip interactions
(``prompt protons'').  Obviously, any additional prompt proton emission would
have the effect to lower the relative neutrino flux consistent with the upper
bound.  However, for the most interesting sources, i.e, GRBs and AGN jets,
the ejection of prompt protons can be expected to be strongly suppressed due
to the rapid expansion of the emission region and the implied large adiabatic
losses. This will allow in the future to use measured neutrino fluxes, or
experimental limits thereof, as a limit to the contribution of these sources
to the cosmic ray flux. If, for example, GRBs produce the observed UHECR flux
and if the global GRB emissivity follows a similar evolution as star
formation and AGNs, bursts would have to emit a neutrino flux on the level of
their optically thin bound, which is roughly two orders of magnitude above
the original prediction \cite{WB97prl}.  This is independent of the total
fractional burst energy converted by the protons in their interactions.  A
factor of ${\sim} 3{-}6$ is due to evolution, a factor of ${\sim}10$ due to
the limited efficiency of $p\to n$ conversion \cite{RM98hunt}, and an
additional factor ${\sim} 5$ due to an accurate treatment of photoproduction
yields in the hard photon fields typical for GRBs (note that the yield
factors used in our figures correspond to AGN-like target photon spectra).
In this case, the gamma rays would also account for most of the EGRB, and one
could observe strong TeV bursts from nearby sources
\cite{Tot98apj2,Tot99app}.  Currently, only AGN jets are known to contribute
to the EGRB in a major way and this has motivated the original model A of
Mannheim~\cite{Man95app} (which is exactly on the level of the optically thin
bound as a consequence of the assumption that AGN jets produce the observed
UHECRs). We note that the neutrino flux due to nucleons escaping the AGN jets
and diffusing through the surrounding AGN host galaxies is difficult to
assess and limited by the EGRB only \cite{Man95app}, although this neutrino flux
is the one most relevant in the $1{-}100\TeV$ range.

Generally, the relevance of a bound on optically thin sources should not be
overestimated.  A large number of extragalactic neutrino sources could be
opaque to the emission of UHE cosmic rays producing a neutrino flux well
above this bound.  For the particularly important class of AGN jets, we have
therefore developed an upper bound to their total neutrino flux using
constraints on their gamma-ray and neutron opacity set by present gamma ray
data.  This {\em maximum} contribution of AGN jets to the extragalactic
diffuse neutrino flux is up to an order of magnitude higher than our
optically thin bound in the energy range between $10^7\GeV$ and
$10^9\GeV$. Additionally, we have discussed that for energies below $\sim
10^8\GeV$ an influence of magnetic fields in the radio lobes of active
galaxies or on larger scales cannot be excluded. For example, the model of
Halzen \& Zas \cite{HZ97apj} predicts a neutrino flux about one order of
magnitude above our upper limit for AGN at $E_\nu\sim 10^8\GeV$, which is
certainly extreme, but still cannot be considered as completely ruled out,
because at this energy the flux of neutrons after turning into cosmic ray
protons may be diminished by an additional factor ${\sim} 10$ due to
adiabatic losses in expanding radio halos. At present, the most general bound
to EGRET detected AGN is given by their contribution to the EGRB.  The most
extreme class of neutrino sources opaque to the emission of UHECRs could be
AGN jets which have {\em not} been detected in gamma rays. Prime candidates
are the GPS and CSS quasars mentioned in Sect.\,\ref{sect:pnu_blazar}, which
together make up about 40\% of the bright extragalactic radio sources.

Finally, we wish to clarify that our result is not in conflict with, but
complementary to, the upper limit previously obtained by Waxman and Bahcall
\cite{WB99prd}. Their result applies to photohadronic sources with a
particular spectral shape which are transparent to the emission of UHE cosmic
rays, such as may be representative of low-luminosity BL Lacertae objects
(like Mrk501) or GRBs. Our result is more general, and therefore less
restrictive, indicating that there may be other classes of sources, such as
quasars, with a different spectral shape, and/or which are opaque to the
emission of UHE cosmic rays, which can produce a higher neutrino flux than
the source classes considered by Waxman and Bahcall. We also confirm the
claim by Waxman and Bahcall that large scale magnetic fields are unlikely to
have an effect for cosmic ray propagation at $\sim 10^{10}\GeV$, but we
additionally point out that magnetic field effects cannot be disregarded at
lower cosmic ray energies. We show that hadronic processes in 
AGN which are optically thick to the emission of UHE cosmic rays {\em could}
produce the extragalactic gamma ray background according to present
observational constraints. However, the limiting model for EGRET-detected AGN
presented in this paper is mainly determined by the current instrumental limits
of gamma-ray astronomy --- future observations in the GeV-to-TeV energy range
may impose stricter bounds, and therefore also limit the possible hadronic
contribution to the EGRB. Together with an independent estimate on the total
contribution of AGN jets to the EGRB, this may allow one to constrain the
overall cosmic ray content of AGN jets in the near future.

\vspace{-2ex}
\section*{Acknowledgements}
\vspace{-2ex}

RJP is supported by the Australian Research Council.  JPR acknowledges
support by the EU-TMR network Astro-Plasma\,Physics, under contract number
ERBFMRX-CT98-0168.  KM acknowledges support through a Heisenberg-Fellowship  
granted by the Deutsche Forschungsgemeinschaft. We thank Torsten Ensslin and 
the referees for helpful comments, and Anita M{\"u}cke for reading the
manuscript.

\ifonecol\raggedright\fi
\bibliographystyle{prsty}
\bibliography{agn,grb,neutrino,uhecr,physics}


\end{document}